\documentclass[11pt,a4paper]{article}
\usepackage{jheppub}
\usepackage{feynmp}
\usepackage[retainorgcmds]{IEEEtrantools}
\usepackage{multirow}
\usepackage{rotfloat}

\usepackage{color}
\usepackage[usenames,dvipsnames]{xcolor}

\title{Renormalisation group invariants and sum rules: \\ 
       fast diagnostic tools for probing high-scale physics}
\author[a]{Jamil Hetzel,}
\author[b]{Wim Beenakker}

\affiliation[a]{Institut f\"ur Theoretische Physik, Universit\"at Heidelberg,\\
Philosophenweg 16, D-69120 Heidelberg, Germany}
\affiliation[b]{Theoretical High Energy Physics, IMAPP, Faculty of Science, 
Mailbox 79, \\ Radboud University Nijmegen, P.O. Box 9010, NL-6500 GL Nijmegen,
The Netherlands}

\emailAdd{hetzel@thphys.uni-heidelberg.de}
\emailAdd{W.Beenakker@science.ru.nl}

\abstract{A method is described to probe high-scale physics in lower-energy 
          experiments by employing sum rules in terms of renormalisation group 
          invariants. The method is worked out in detail for the study of 
          supersymmetry-breaking mechanisms in the context of the Minimal 
          Supersymmetric Standard Model. To this end sum rules are constructed 
          that test either specific models of supersymmetry breaking or 
          general properties of the physics that underlies supersymmetry 
          breaking, such as unifications and flavour-universality.}

\keywords{Renormalization Group, Sum Rules, Beyond Standard Model, 
          Supersymmetry Breaking}

\numberwithin{equation}{section}

\begin{document}
\maketitle\marginpar{}

\renewcommand{\d}{\mathrm{d}}
\newcommand{\lag}{\mathcal{L}}
\newcommand{\ord}[1]{\mathcal{O}\left(#1\right)}
\newcommand{\msq}{m_{\widetilde{Q}}^2}
\newcommand{\msqa}{m_{\widetilde{Q}_1}^2}
\newcommand{\msqc}{m_{\widetilde{Q}_3}^2}
\newcommand{\msu}{m_{\widetilde{\bar{u}}}^2}
\newcommand{\msua}{m_{\widetilde{\bar{u}}_1}^2}
\newcommand{\msuc}{m_{\widetilde{\bar{u}}_3}^2}
\newcommand{\msd}{m_{\widetilde{\bar{d}}}^2}
\newcommand{\msda}{m_{\widetilde{\bar{d}}_1}^2}
\newcommand{\msdc}{m_{\widetilde{\bar{d}}_3}^2}
\newcommand{\msl}{m_{\widetilde{L}}^2}
\newcommand{\msla}{m_{\widetilde{L}_1}^2}
\newcommand{\mslc}{m_{\widetilde{L}_3}^2}
\newcommand{\mse}{m_{\widetilde{\bar{e}}}^2}
\newcommand{\msea}{m_{\widetilde{\bar{e}}_1}^2}
\newcommand{\msec}{m_{\widetilde{\bar{e}}_3}^2}
\newcommand{\mhu}{m_{H_u}^2}
\newcommand{\mhd}{m_{H_d}^2}

\section{Introduction}\label{s:introduction}

Although the Standard Model of electroweak interactions has worked pretty well 
so far, it has a number of shortcomings. For example, it is difficult to 
explain why the Higgs mass is much smaller than the Planck scale (known as the 
``hierarchy problem''), a description of gravity is lacking and there is no 
good candidate for dark matter. Moreover, given the fact that the Higgs mass is
either relatively light or rather heavy, i.e.~$122$ GeV $<M_H<127$~GeV or $M_H>600$~GeV 
at 95\% confidence level with a possible Higgs discovery at a mass near 
$125$~GeV \cite{LEPHiggs,ATLASHiggs,CMSHiggs,ATLASdiscovery,CMSdiscovery}, it is difficult to 
guarantee the stability of the Standard Model Higgs mechanism up to the Planck 
scale \cite{chimneyplot}. This might hint at the possibility that there is a 
scale of new physics between the electroweak scale and the Planck scale.

If signs of such new physics are observed at the upcoming runs of the Large
Hadron Collider (LHC), the main theoretical challenge will be to unravel the
underlying theory. Since the underlying physics can reside at energies that
largely exceed the reach of the LHC, this is not going to be a simple task. 
The standard strategies for addressing this issue make use of renormalisation 
group techniques, linking the ``low-energy'' physics observed at present-day 
experiments to high-energy parameters at the energy scale where the underlying 
theory is formulated. The most widely studied approach is the top-down method,
where one starts by choosing a specific new-physics model. Subsequently, the 
high-scale model parameters are evolved down to the collider scale and  
predictions can be made about the way the model will manifest itself
phenomenologically in ongoing experiments, allowing a confrontation between 
theory and experiment. This approach is great for identifying where one should
look for signs of new physics, but it is not really well-suited for deriving 
conclusive statements about the underlying model. Alternatively a bottom-up 
method can be used, where one is guided by experimental data. One starts by 
adopting a rather general phenomenological framework for describing the physics
beyond the Standard Model, such as a supersymmetric extension of the Standard 
Model. Within this phenomenological context, the data are converted into 
running parameters at the collider scale. Subsequently, the running parameters 
are evolved up towards the scale where the underlying physics is presumably 
residing, allowing the high-scale parameters to be confronted with specific 
predictions from new-physics models. This method is better-suited for getting 
information on the underlying model. However, since the renormalisation group 
evolution is a numerical procedure, it can lead to an increase of the error from one scale to another. 
Moreover, the fact that we do not know the scale of new physics 
can lead to misinterpretation of the data. A more detailed discussion of the 
advantages and drawbacks of the top-down and bottom-up methods will be given in
section~2.

Recently, an elegant third approach to probing high-scale physics has gained 
some interest \cite{Demir,Carena1,Carena2,Jaeckel}. This approach is based on 
the same philosophy as the bottom-up method, but instead of using the full set
of running parameters it makes clever use of so-called renormalisation group 
invariants. These are combinations of running parameters chosen in such a way 
that they do not evolve with energy. If we measure their values at the collider
scale, we will immediately know their values at the threshold of new physics. 
This fact allows one to probe physics at high energy scales without having to 
evolve all parameters. Subsequently, the renormalisation group invariants can 
be combined into sum rules that test the underlying physics. Up to now, sum 
rules have been constructed for testing specific models. We advocate to employ 
sum rules in a more model-independent way, by using them as fast diagnostic 
tools to test generic properties that are common to new-physics models, such as
unification and universality properties. If a certain property is realised in 
Nature, all corresponding sum rules must be satisfied. So, the main strength of
invariant sum rules is their falsifying power. 

In order to give an idea of how these renormalisation-group-invariant sum rules
work in practice, we work them out in detail for the study of 
supersymmetry-breaking mechanisms in the context of a phenomenological version 
of the Minimal Supersymmetric Standard Model. This model is one of the prime 
beyond-the-Standard-Model frameworks to be tested at the LHC, since it offers 
solutions to several of the problems that plague the Standard Model. 

The structure of the paper is as follows. In section~2 we describe the
concept of effective field theories and discuss renormalisation
group techniques. In section~3 we give the salient details of the Minimal 
Supersymmetric Standard Model and a few popular supersymmetry-breaking 
mechanisms. Subsequently, the one-loop renormalisation group invariants
are listed for a phenomenological version of the model. In section~4 we give 
a detailed discussion of the model-independent and model-specific sum rules 
that can be used for studying the supersymmetry-breaking mechanisms. We will 
conclude in section~5. 

\begin{fmffile}{efts}
\section{Effective Lagrangians and renormalisation group equations}
\label{s:efts}
In a renormalisable quantum field theory, all divergences can be dealt with by 
redefining the masses and couplings, which become scale dependent. On the other 
hand, if the Lagrangian contains operators with dimension greater than four, an infinite 
number of counterterms is needed to remove all divergences and all predictive 
power is lost. Hence, the prevalent view on renormalisation used to be that a sensible 
theory describing Nature could only contain renormalisable interactions. This 
view has changed since Wilson's work on the Renormalisation Group (RG) \cite{wilsonkogut,wilson}. 
The use of a momentum cutoff $\Lambda$ is now considered to have a physical 
meaning: $\Lambda$ is the scale at which new physics becomes relevant. For processes at 
energies greater than $\Lambda$, the theory is not valid anymore and should be replaced by a 
more fundamental theory. This is the motivation for using effective field theories (EFTs).

\subsection{Effective field theories}\label{s:efts-efts}
Physics problems usually involve widely separated energy scales, which allows us to 
study low-energy dynamics without needing to know the details of the high-energy interactions.
Effective field theories (see e.g.\ \cite{eft}) are the theoretical tool to 
describe low-energy physics, where `low' means low with respect to some energy 
scale $\Lambda$. An EFT only takes into account states with mass $m\ll\Lambda$;
heavier excitations with $m\gg\Lambda$ are integrated out from the action.
The information about the heavy states is then contained in the couplings of 
the low-energy theory: we get non-renormalisable interactions among the light 
states, organised as an expansion in powers of energy/$\Lambda$.

An effective field theory is characterised by some effective Lagrangian:
\begin{equation}
\lag = \sum_ic_iO_i,
\end{equation}
where the $O_i$ are operators constructed from the light fields and the $c_i$ 
are couplings containing information on any heavy degrees of freedom. Since the
Lagrangian has dimension~4, dimensional analysis yields:
\begin{equation}
[O_i] \equiv d_i \qquad\Rightarrow\qquad c_i \sim \frac{1}{\Lambda^{d_i-4}},
\end{equation}
where $\Lambda$ is some characteristic heavy scale of the system.
At low energies, the behaviour of these operators is determined by their 
dimension:
\begin{itemize}
\item Operators with $d_i<4$ are called \emph{relevant}, since they give rise 
      to effects that become large at low energies.
\item Operators with $d_i>4$ are called \emph{irrelevant}: at energy scales $E$
      their effects are suppressed by powers of $E/\Lambda$, making them small 
      at low energies. These are non-renormalisable operators that contain 
      information about the underlying dynamics at higher scales.
\item Operators with $d_i=4$ are called \emph{marginal}, because they are 
      equally important at all energy scales.
\end{itemize}
This explains why we are able to include non-renormalisable operators in an EFT
without spoiling its predictive power: at low energies $E$, their effects can
be either neglected or incorporated as perturbations in powers of $E/\Lambda$.
At high energies, it is more appropriate to use a different EFT. Thus at 
sufficiently low energies, an EFT automatically contains only renormalisable 
operators.

\subsection{Matching}\label{s:efts-matching}
Suppose we have two EFTs: one that includes a heavy particle and one where its 
effects are included in the form of higher-dimensional operators, suppressed by
inverse powers of the heavy particle mass $M$. Since physics around the mass 
scale $M$ should not depend on our choice of theory, both EFTs should yield the
same physical predictions. Hence they are related by the \emph{matching 
condition}: at the threshold $\mu=M$, the two EFTs should give rise to the same
$S$-matrix elements for light-particle scattering. This leads to relations 
between the parameters of the high-energy EFT (the one we use above threshold) 
and those of the low-energy EFT (the one we use below threshold). In other 
words, the matching conditions encode the effects of the heavy field into the 
low-energy EFT parameters.

As an example, consider the beta decay of a neutron. In the Standard Model, 
this decay is mediated by a $W$ boson with mass $M_W$. At energies $\mu\ll M_W$, 
there is not enough energy available to produce a physical $W$ boson. Hence we might as well 
integrate out the $W$ field from the action. In the resulting low-energy EFT, the $W$ exchange reduces to a non-renormalisable 
contact interaction. Matching the two EFTs at $\mu=M_W$ yields the formula for the Fermi coupling constant
\begin{equation}
G_F = \frac{\sqrt{2}}{8}\frac{g_2^2}{M_W^2},
\end{equation}
where $g_2$ is the weak coupling constant. Note that although the $W$ field is 
not included in the low-energy EFT, its `fingerprints' (namely its coupling constant $g_2$ and mass $M_W$) 
are still present in the low-energy coupling $G_F$. Also note that the irrelevant operator 
corresponding to the contact interaction is indeed suppressed by powers of $M_W$, as mentioned 
in section~\ref{s:efts-efts}.

\subsection{How to look for clues about new physics}\label{s:efts-travelling}
In the process of renormalising a theory, we redefine the masses and couplings 
by having them depend on a reference scale $\mu$. This $\mu$-dependence is given 
by the renormalisation group equations, which can be calculated from the condition 
that anything observable should be independent of $\mu$. These equations depend on 
the loop order of the calculations and on the particle content of the theory.

In the framework of effective field theories, physics is described by a chain of EFTs.
The RG equations and matching conditions for these EFTs allow us to evolve a 
theory between different energy scales.
Suppose that we have an EFT describing physics at some (high) energy scale 
$\mu$. If we want to describe physics at a lower energy scale, we have to evolve down 
the running parameters using the RG equations of this EFT.
Whenever we reach a particle threshold, we must switch to an EFT that does not contain 
the corresponding field. The appropriate matching conditions at the particle threshold yield the parameters 
of the low-energy EFT. From there we can continue to evolve these parameters, now using the 
RG equations of the low-energy EFT.

\begin{figure}[t]
\begin{center}
{\setlength{\unitlength}{1.35pt}
\begin{picture}(290,100)(-50,-70)
  \put(0,-70){\vector(0,1){100}}
  \put(-8,24){$\mu$}
  \put(-3,0){\line(1,0){6}}
  \multiput(0,0)(2,0){5}{\line(1,0){1}}
  \multiput(57,0)(2,0){38}{\line(1,0){1}}
  \multiput(229,0)(2,0){4}{\line(1,0){1}}
  \put(-50,2){new physics}
  \put(-45,-6){threshold}
  \put(20,2){evolved}
  \put(13,-6){parameters}
  \put(9,-9){\line(1,0){48}}
  \put(9,-9){\line(0,1){18}}
  \put(57,9){\line(-1,0){48}}
  \put(57,9){\line(0,-1){18}}
  \put(66,-12){compare theory}
  \put(65,-20){with experiment}
  \thicklines
  \put(57,-5){\vector(1,0){75}}
  \put(132,-5){\vector(-1,0){75}}
  \thinlines
  \put(136,2){RG boundary conditions}
  \put(156,-6){from matching}
  \put(132,-9){\line(1,0){97}}
  \put(132,-9){\line(0,1){18}}
  \put(229,9){\line(-1,0){97}}
  \put(229,9){\line(0,-1){18}}
  \put(134,25){more fundamental theory}
  \put(151,-27){low-energy EFT}
  \put(113,-70){experimentally accessible regime}
  \multiput(57,-63)(3,0){58}{\line(1,3){9}}
  \put(231,-63){\line(1,3){6}}
  \put(234,-63){\line(1,3){3}}
  \multiput(6,-72)(3,0){34}{\line(1,3){3}}
  \multiput(3,-72)(0,9){3}{\line(1,3){6}}
  \put(3,-45){\line(1,3){3}}
  \multiput(12,-36)(3,0){17}{\line(-1,-3){3}}
  \put(63,-36){\line(-1,-3){6}}
  \put(16,-52){measured}
  \put(13,-60){parameters}
  \put(9,-63){\line(1,0){48}}
  \put(9,-63){\line(0,1){18}}
  \put(57,-45){\line(-1,0){48}}
  \put(57,-45){\line(0,-1){18}}
  \thicklines
  \put(33,-45){\vector(0,1){34}}
  \thinlines
  \put(36,-27){RG}
\end{picture}}
\end{center}
\caption{Scheme for studying physics at scales beyond experimental access. 
         The running couplings are measured at a scale where the low-energy EFT
         is applicable. Using the RG equations, they are evolved towards the 
         threshold where new fields presumably enter the theory. Then they can 
         be compared with the matching conditions predicted by the more 
         fundamental theory.}\label{f:EFTtravelling}
\end{figure}
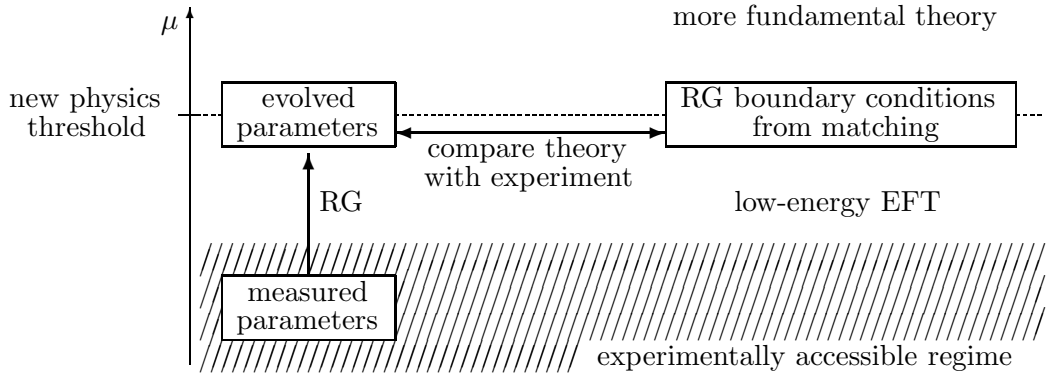

From this point of view, the Standard Model is only a low-energy EFT  
of Nature. The shortcomings of the Standard Model hint at the 
existence of a more fundamental theory. Even if that more fundamental EFT is 
appropriate only at energies beyond experimental access, the idea of a chain of 
EFTs certainly helps us study that more fundamental theory (figure~\ref{f:EFTtravelling}). We could measure 
the running parameters at a low scale $\mu$ and then evolve them upwards.
At the threshold of the more fundamental theory, the matching conditions act as
boundary conditions for the renormalisation group. Hence, by comparing our 
evolved masses and couplings with the predicted matching conditions, we can get
information on the high-energy theory.

\subsection{How to probe the high scale}\label{s:efts-probehighscale}
\subsubsection{Top-down method}\label{s:efts-topdown}
The literature offers various approaches to using the renormalisation group to 
extract information about high-scale matching conditions. In the context of 
supersymmetry, the most widely studied one is the top-down method. It is called this way because a 
top-down study is started from the high scale and the theory is evolved down to
the collider scale (say $\ord{\text{1 TeV}}$). One starts by choosing a 
specific high-scale new-physics model with a limited number of free parameters and proceeds as follows: 
\begin{itemize}
\item Pick a point in the high-scale parameter space of the model and translate this into 
      values of the running parameters at the new-physics threshold (masses, gauge and 
      Yukawa couplings, etc.). 
\item Evolve the running parameters down to the collider scale using the 
      renormalisation group equations.
\item Using the resulting parameter values, calculate the 
      relevant branching ratios and cross sections. 
\item Compare the results to experimental data and extract constraints on the 
      high-scale parameter space of the model. 
\end{itemize}
The top-down method is suitable for making general phenomenological 
predictions. For example, it is used to find collider signatures that are 
characteristic for supersymmetry. However, for the purpose of testing a 
new-physics model, this method has some serious limitations:
\begin{itemize}
\item With the top-down method one can only determine the regions in the 
      high-scale parameter space of the model that are consistent with the data. If only 
      small portions of the total parameter space seem phenomenologically 
      viable, one might conclude that the model is neither likely to be correct
      nor natural. However, it seems unlikely that we can strictly exclude a 
      model this way.
\item In this approach the confrontation between theory and experiment is performed on the basis of individual models.
      Given the fact that there are many high-scale models on the market and most of them can also
      be tweaked in one way or another, it seems unlikely that we can find the right
      model this way. 
\item Scanning the entire high-scale parameter space of even a single model is very time-consuming. To scan it 
      properly, one ought to use a reasonably fine grid and check each point 
      separately. But the parameter space is usually too big to perform a full 
      detector simulation for each point. For general predictions of 
      supersymmetric phenomenology, for example, one usually resorts to using a
      limited set of benchmark points (see e.g.\ \cite{benchmarkpoints}), 
      because many points in parameter space have a very similar phenomenology.
      However, for the purpose of excluding a certain model, this is no 
      satisfying solution.
\item Fitting the numerical predictions to the experimental data becomes much 
      more difficult as the number of high-scale model parameters is increased. Therefore 
      one always limits oneself to a model with few parameters. But there is no
      reason to think that Nature would restrict itself to only a few 
      parameters in the EFT beyond the next threshold.
\end{itemize}

\subsubsection{Bottom-up method}\label{s:efts-bottomup}
The bottom-up method is an alternative to the top-down method. 
It works by evolving the theory upwards from the collider 
scale to the new-physics threshold. A bottom-up analysis consists of the 
following steps:
\begin{itemize}
\item Convert experimental data into the running parameters at the collider 
      scale.
\item Using the renormalisation group equations pertaining to the low-energy 
      EFT (e.g.\ the Standard Model above the electroweak scale, or the Minimal
      Supersymmetric Standard Model 
      above the supersymmetric mass scale), evolve these running parameters 
      towards the scale where new physics presumably comes into play.
\item Analyse the structure of the high-scale parameters: do they fit the 
      matching conditions predicted by any new-physics model?
\end{itemize}
This method seems more suitable for excluding new-physics models than the 
top-down method. Also, there is no practical need to only consider high-scale new-physics models with 
a limited number of free parameters. Furthermore, no time-consuming scanning of that parameter space 
is involved. However, the bottom-up method presents challenges of its own:
\begin{itemize}
\item The running parameters at the collider scale will come with experimental 
      errors. To determine the uncertainty in these parameters at a higher 
      scale, we also have to evolve the error bars. These may become larger 
      while numerically performing the renormalisation group evolution, which 
      could make it difficult to tell for example whether certain parameters 
      unify or not.
\item We do not know the value of the high scale that should be taken as the 
      new-physics threshold; this scale has to be guessed. In practice, one 
      might evolve the running parameters until some of them unify and take the
      corresponding scale as the new-physics threshold. But a unification scale
      does not necessarily correspond to a threshold.\footnote{This occurs for 
      example in a supersymmetry-breaking model called Mirage Mediation, see 
      section~\ref{s:susy}.} Also, there may be an intermediate new-physics 
      threshold even though no unification occurs there. In both cases, we 
      would extract incorrect boundary conditions for matching with the 
      underlying high-energy EFT.
\item Because the RG equations are coupled, all running parameters must be 
      known. Hence if we fail to measure one mass or coupling, 
      the bottom-up method cannot be used except for subsets of parameters 
      whose RG equations contain only the parameters from that subset.
\end{itemize}

\subsubsection{Renormalisation group invariants}\label{s:efts-rgis}
Recently, a third approach to probing the high scale has gained some interest 
\cite{Demir,Carena1,Carena2,Jaeckel}. This approach makes clever use of 
renormalisation group invariants (RGIs). These are combinations of running 
parameters chosen in such a way that they are independent of the 
renormalisation scale $\mu$. A well-known example of an RGI is the following 
combination of the strong coupling $g_3$, the weak coupling $g_2$ as well as 
the scaled hypercharge coupling $g_1=g'\sqrt{5/3}$:
\begin{equation}
I_g \equiv (b_2-b_3)g_1^{-2} + (b_3-b_1)g_2^{-2} + (b_1-b_2)g_3^{-2},
\end{equation}
where the gauge couplings satisfy the following renormalisation group equations
at one loop:
\begin{equation}
16\pi^2\frac{\d g_a}{\d t} = b_ag_a^3 \qquad(a=1,2,3).
\label{eq:gaugecouplingRGEs}
\end{equation}
Here $t\equiv\ln{(\mu/\mu_0)}$, where $\mu_0$ is a reference scale that makes 
the argument of the logarithm dimensionless; its value is arbitrary since it 
drops out of the RG equations. The RG coefficients $b_a$ are constants 
depending on the particle content of the model. For the Standard Model they are
$b_a=(\frac{41}{10},-\frac{19}{6},-7)$ for $a=1,2,3$, whereas for the Minimal 
Supersymmetric Standard Model (MSSM) they are $b_a=(\frac{33}{5},1,-3)$. It is 
easily checked that $\d I_g/\d t=0$. Note that $I_g$ is not exactly RG 
invariant, since we used the one-loop RG equations to construct it. We will 
come back to this issue in section~\ref{s:results}.

A crucial property of RGIs is that if we measure their values at the collider 
scale, we will immediately know their values at the threshold of new physics.
This fact allows us to probe physics at high energy scales without having to 
evolve all parameters. For example, $I_g$ can be used to test whether the gauge
couplings unify. Note that if the gauge couplings have a universal value 
$g_{_U}$ at some energy scale, then $I_g$ will vanish. Since $I_g$ is an RGI, 
it will consequently vanish at every scale where the renormalisation group 
equations \eqref{eq:gaugecouplingRGEs} are valid. Hence, if we measure the 
gauge couplings at one scale, we can perform a quick diagnostic check to test 
whether gauge-coupling unification occurs within the context of a specific EFT.
Also note that this consistency check is independent of the value of the scale 
where the gauge couplings unify.
\begin{figure}[t]
\begin{center}
  \includegraphics[width=.5\textwidth]{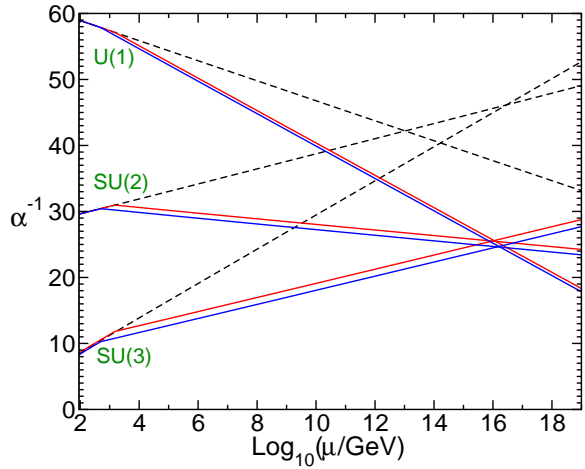}
\end{center}
\caption{Two-loop renormalisation group evolution of the inverse gauge 
         couplings squared $\alpha_a^{-1}(\mu) = 4\pi g_a^{-2}(\mu)$ for 
         $a=1,2,3$ in the Standard Model (dashed lines) and the Minimal 
         Supersymmetric Standard Model (solid lines). For the latter, the 
         supersymmetric particle thresholds are varied between 500 GeV and 
         1.5 TeV, and $\alpha_3(M_Z)$ is varied between 0.117 and 0.121. 
         Figure taken from \protect\cite{martin}.}\label{f:gaugeunification}
\end{figure}

To illustrate this, we perform this check explicitly for the Standard Model, 
using the measured couplings at $\mu=M_Z$ in the $\overline{MS}$ scheme.
The gauge couplings can be obtained from the measured quantities 
$\alpha_3^{-1}(M_Z) = 8.45 \pm 0.05$,
$\alpha^{-1}(M_Z) = 127.916 \pm 0.015$ and
$\sin^2\theta_W(M_Z) = 0.23131 \pm 0.00007$ \cite{PDG}.
Here $\alpha_a^{-1} \equiv 4\pi g_a^{-2}$, $\alpha$ is the fine structure 
constant and $\theta_W$ is the weak mixing angle.
The latter two are related to $\alpha_2$ and $\alpha_1$ by the relations
\begin{align}
\alpha_2^{-1}(M_Z) &= \sin^2\theta_W(M_Z) \alpha^{-1}(M_Z) = 29.588 \pm 0.010, 
\notag\\
\alpha_1^{-1}(M_Z) &= \frac35\cos^2\theta_W(M_Z) \alpha^{-1}(M_Z) = 58.997 \pm 0.009.
\end{align}
Using these values we find $I_g^\text{SM} = -3.252 \pm 0.030$, which lies many 
standard deviations from~$0$. Hence we find no compatibility with 
gauge-coupling unification within the Standard Model; this can be confirmed 
using the bottom-up method (figure~\ref{f:gaugeunification}).

Similarly, we can make an estimate for $I_g$ in the MSSM by taking the above 
values of $\alpha_a^{-1}(M_Z)$.\footnote{In order to determine the actual value
   of $I_g^\text{MSSM}$ we ought to use the values of $\alpha_a^{-1}$ at the 
   scale where the MSSM becomes valid, i.e.\ the highest supersymmetric 
   particle threshold. This scale will be somewhat higher than $M_Z$. As can be
   seen using the bottom-up method (figure~\ref{f:gaugeunification}), this 
   barely changes the prediction of gauge-coupling unification: only the value 
   of the unification scale might change. Hence we should get a good estimate 
   of $I_g^\text{MSSM}$ using the values of $\alpha_a^{-1}(M_Z)$.}
We find $I_g^\text{MSSM}=-0.059\pm 0.024$, which is close to zero.
Hence the MSSM might allow for gauge-coupling unification, depending on the 
actual values of the supersymmetric particle thresholds. This can be confirmed 
using the bottom-up method (figure~\ref{f:gaugeunification}): the 
gauge couplings are observed to unify, even when we use the two-loop RG equations. 

To summarise, RGIs provide a fast diagnostic tool for probing matching 
conditions at high energy scales. They circumvent the need to evolve the 
running parameters numerically; we do not even need to know exactly at which 
energy scale new physics arises.
\end{fmffile}

\begin{fmffile}{susy}
\section{Using RGIs and sum rules to study supersymmetry breaking}
\label{s:susy}
In order to give an idea of how the RGI method works in practice, we work it 
out explicitly for models of supersymmetry breaking. To this end, we first give
a short description of the salient details of the minimal supersymmetric 
extension of the Standard Model. This is followed by a discussion of the 
necessity to break supersymmetry and the ways to achieve this. For an extensive
introduction to the subject the interested reader is referred to 
Refs.~\cite{martin,aitchison}.

\subsection{Supersymmetry}\label{s:susy-susy}
Although the Standard Model has worked pretty well so far, it has a number of 
shortcomings which we mentioned in the introduction. 
Supersymmetry offers possible solutions to these problems 
\cite{martin,aitchison}. It is a symmetry between fermions and bosons that 
would complete the list of possible spacetime symmetries \cite{CMtheorem,HLSextension}. 
The single-particle states of a 
supersymmetric theory fall into irreducible representations of the 
supersymmetry algebra, called supermultiplets. These contain an equal number of
bosonic and fermionic degrees of freedom. The bosons and fermions in a 
supermultiplet are called \emph{superpartners} of each other.
A minimal supersymmetric extension of the Standard Model contains two types of 
supermultiplets: \emph{chiral (matter) supermultiplets}, which consist of a 
two-component Weyl spinor and a complex scalar field, and \emph{vector (gauge) 
supermultiplets}, which consist of a spin-1 gauge-boson field and a spin-1/2 
Majorana spinor, called the \emph{gaugino} field. The supersymmetry generators 
$Q$, $Q^\dagger$ commute with the mass-squared operator $P^2$ and the 
generators of gauge transformations, so superpartners have the same mass and 
gauge quantum numbers. In view of the quantum-number structure of the Standard
Model, this implies that a supersymmetric extension of the Standard Model 
introduces at least one new supersymmetric particle (or \emph{sparticle} for 
short) for each Standard Model particle. 

\subsection{The Minimal Supersymmetric Standard Model}\label{s:susy-mssm}
The Minimal Supersymmetric Standard Model (MSSM) is defined to be a 
supersymmetric extension of the Standard Model with minimal particle content and a 
new conserved quantum number called $R$-parity \cite{Fayet}. 
The MSSM particle content and nomenclature is listed in 
tables~\ref{t:mssmchiralcontent}--\ref{t:mssmgaugecontent}.
The only new parameter with respect to the Standard Model is a supersymmetry-preserving 
Higgs-mixing parameter $\mu$.

\begin{table}[t]
\begin{center}
{\renewcommand{\arraystretch}{1.5}
\begin{tabular}{|c|c|c|c|c|}
\hline
\multirow{2}{3cm}{\begin{center}Name\end{center}}	&	\multirow{2}{1.5cm}{\begin{center}Symbol\end{center}}	&	\multirow{2}{1.5cm}{\begin{center}Spin 0\end{center}}	&	\multirow{2}{1.5cm}{\begin{center}Spin 1/2\end{center}}	&	Gauge-group	\\
	&	&	&	& representation	\\\hline\hline
\multirow{3}{3cm}{\begin{center}squarks \& quarks (3 generations)\end{center}}	&	$Q$	&	$\widetilde{Q}=(\widetilde{u}_L \,,\, \widetilde{d}_L)$	&	$(u_L \,,\, d_L)$	&	$(\mathbf{3},\mathbf{2},\frac16)$	\\
		&	$\bar{u}$	&	$\widetilde{u}_R^{\,*}$	&	$u_R^c$	&	$(\mathbf{\bar{3}},\mathbf{1},-\frac23)$	\\
		&	$\bar{d}$	&	$\widetilde{d}_R^{\,*}$	&	$d_R^c$	&	$(\mathbf{\bar{3}},\mathbf{1},\frac13)$	\\\hline
sleptons \& leptons	&	$L$	&	$\widetilde{L}=(\widetilde{\nu} \,,\, \widetilde{e}_L)$	&	$(\nu \,,\, e_L)$	&	$(\mathbf{1},\mathbf{2},-\frac12)$	\\
(3 generations)		&	$\bar{e}$	&	$\widetilde{e}_R^{\,*}$	&	$e_R^c$	&	$(\mathbf{1},\mathbf{1},1)$	\\\hline
\multirow{2}{3cm}{\begin{center}Higgs \& higgsinos\end{center}}	&	$H_u$	&	$(H_u^+ \,,\, H_u^0)$	&	$(\widetilde{H}_u^+ \,,\, \widetilde{H}_u^0)$	&	$(\mathbf{1},\mathbf{2},\frac12)$ \\
	&	$H_d$	&	$(H_d^0 \;,\, H_d^-)$	&	$(\widetilde{H}_d^0 \;,\, \widetilde{H}_d^-)$	&	$(\mathbf{1},\mathbf{2},-\frac12)$ \\\hline
\end{tabular}}
\caption{Chiral supermultiplet content of the Minimal Supersymmetric Standard 
         Model and the corresponding representations of the gauge group 
         $SU(3)_C\times SU(2)_L\times U(1)_Y$, with superpartners indicated by 
         a tilde. Note that we need an additional Higgs doublet compared to the
         Standard Model and that right-handed modes are charge conjugated in 
         order to bring them into left-handed form.}\label{t:mssmchiralcontent}
\end{center}
\end{table}
\begin{table}[t]
\begin{center}
{\renewcommand{\arraystretch}{1.5}
\begin{tabular}{|c|c|c|c|}
\hline
\multirow{2}{3cm}{\begin{center}Names\end{center}}	&	\multirow{2}{1.5cm}{\begin{center}Spin 1/2\end{center}}	&	\multirow{2}{1.5cm}{\begin{center}Spin 1\end{center}}	&	Gauge-group	\\
	&	&	&	representation	\\\hline\hline
gluino \& gluon		&	$\widetilde{g}$	&	$g$	&	$(\mathbf{8},\mathbf{1},0)$	\\\hline
winos \& $W$ bosons	&	$\widetilde{W}^1 \quad \widetilde{W}^2 \quad \widetilde{W}^3$	&	$W^1 \quad W^2 \quad W^3$	&	$(\mathbf{1},\mathbf{3},0)$	\\\hline
bino \& $B$ boson	&	$\widetilde{B}^0$	&	$B^0$	&	$(\mathbf{1},\mathbf{1},0)$	\\\hline
\end{tabular}}
\caption{Gauge supermultiplet content of the Minimal Supersymmetric Standard 
         Model and the corresponding representations of the gauge group 
         $SU(3)_C\times SU(2)_L\times U(1)_Y$.}\label{t:mssmgaugecontent}
\end{center}
\end{table}

\subsection{Constraints on broken supersymmetry}\label{s:susy-constraints}
If supersymmetry were an exact symmetry of Nature, each sparticle would have 
the same mass as its Standard Model partner and we would have discovered them 
already. Hence, if supersymmetry is a symmetry of Nature, it must be broken 
somehow. The requirement that broken supersymmetry should still solve the 
problems of the Standard Model puts constraints on the possible terms of a 
supersymmetry-breaking Lagrangian.

In order to maintain the solution to the hierarchy problem, we must 
consider \emph{soft supersymmetry breaking}. This means that we only consider 
supersymmetry-violating terms with masses and couplings of \emph{positive} dimension.
By using only relevant operators (see section~\ref{s:efts-efts}) to break supersymmetry, we guarantee that the 
high-scale physics responsible for supersymmetry breaking decouples at low 
energies. It also guarantees that non-supersymmetric corrections to the Higgs 
mass vanish in the limit $m_\text{soft}\rightarrow0$, where $m_\text{soft}$ is 
the largest mass scale associated with the soft parameters.

The most general soft-supersymmetry-breaking Lagrangian in the MSSM, compatible
with gauge invariance and $R$-parity conservation, contains complex gaugino 
masses $M_1$, $M_2$, $M_3$; trilinear couplings $\mathbf{a_u}$, $\mathbf{a_d}$ 
and $\mathbf{a_e}$, which are complex $3\times3$ matrices in family space 
similar to the Yukawa couplings; sfermion mass terms 
$\mathbf{m_Q^2}, \mathbf{m_{\bar{u}}^2}, 
\mathbf{m_{\bar{d}}^2},$ $\mathbf{m_L^2}, \mathbf{m_{\bar{e}}^2}$, which are 
Hermitian $3\times3$ mass matrices in family space; real Higgs masses 
$m_{H_u}^2$ and $m_{H_d}^2$ and a complex supersymmetry-breaking Higgs mixing 
parameter $b$.

Soft supersymmetry breaking introduces 97 new masses, mixing 
angles and phases \cite{manyparameters}. These may lead to predictions for 
flavour-changing neutral current (FCNC) and CP-violating processes that violate 
experimental bounds.
In order to suppress these processes, we 
additionally assume the following relations between the parameters:
\begin{itemize}
\item The soft sfermion masses are flavour diagonal and the first- and 
      second-generation masses are degenerate.
\item There are no sources of CP-violation in the soft-supersymmetry-breaking 
      sector beyond those induced by the Yukawa couplings.
\end{itemize}
We also neglect the first- and second-generation Yukawa and trilinear 
couplings, because they give very small contributions to the evolution of the 
soft-supersymmetry-breaking parameters. These contributions are smaller than 
the two-loop corrections associated with the gauge couplings and 
third-generation Yukawa couplings.

These assumptions resemble the ones that form the basis for the so-called
phenomenological MSSM (pMSSM) \cite{pmssm}. Note, however, that we opt to work 
with the soft-supersym\-metry-breaking parameters in our approach rather than 
the mass eigenstates that are used in the pMSSM.
Under these assumptions, we are left with the following parameters:
\begin{itemize}
\item Twelve real soft scalar masses, which we denote as $\msqa$, $\msqc$, 
$\msua$, $\msuc$, $\msda$, $\msdc$, 
$\msla$, $\mslc$, $\msea$, $\msec$, $\mhu$, $\mhd$, in accordance with the 
notation in table \ref{t:mssmchiralcontent}. The subscripts 1 and 3 refer to 
the first and the third generation respectively.
\item Three real gauge couplings $g_1$, $g_2$, $g_3$.
\item Three real gaugino masses $M_1$, $M_2$, $M_3$.
\item Three real third-generation Yukawa couplings $y_t$, $y_b$, $y_\tau$.
\item Three real soft trilinear couplings $A_t$, $A_b$, $A_\tau$ defined by
\begin{equation}
a_i = A_i y_i \qquad (i=t,b,\tau, \text{ no summation}).
\end{equation}
\item Two real Higgs mixing parameters $\mu$, $b$.
\end{itemize}

One might even go one step further and assume full universality, i.e.~take all
mass matrices proportional to the unit matrix. Such apparently arbitrary 
relations between the soft parameters could make sense from the 
effective-field-theoretical point of view. If supersymmetry is exact in a more 
fundamental EFT than the MSSM, but is broken spontaneously at some high energy 
scale, then the terms in $\lag_\text{soft}$ may arise as effective 
interactions. In that case, the universality relations could arise as matching 
conditions at the threshold where we switch from the more fundamental theory to
the MSSM. Strictly speaking, flavour-universality is lost once the parameters
are evolved down to the electroweak scale, but the numerical impact of this
RG evolution is small \cite{minflav}. Thus the desire for a theory that 
naturally explains supersymmetry breaking forces us to consider spontaneously 
broken supersymmetry.

\subsection{Breaking supersymmetry}\label{s:susy-breaking}
For a spontaneous breakdown of supersymmetry we need a Lagrangian that 
preserves supersymmetry but a vacuum state that breaks it. During the 
construction of a supersymmetric theory, one has to introduce 
\emph{auxiliary fields} for each supermultiplet in order to make the 
supersymmetry algebra close off-shell. These are scalar fields that turn out to
be suitable for breaking supersymmetry (see e.g.\ section 7 of \cite{martin}): 
if some of them acquire a non-zero vacuum expectation value (VEV), 
supersymmetry is broken.

It turns out to be difficult to make this happen using only renormalisable 
interactions at tree level. Therefore, the MSSM soft terms are expected to 
arise radiatively. In radiative supersymmetry-breaking models, supersymmetry is
broken in a \emph{hidden sector}, which contains fields that have no direct 
couplings to the MSSM fields. The latter are said to be in the \emph{visible 
sector}. The two sectors only interact indirectly; the interactions between 
them are responsible for mediating the supersymmetry breakdown from the hidden 
sector to the MSSM. If the mediating interactions are flavour blind, then the 
soft terms of the MSSM will automatically satisfy universality conditions. 
We will discuss several proposals for breaking mechanisms in the sections 
\ref{s:susy-SUGRA}--\ref{s:susy-MM}.

\subsection{Supergravity}\label{s:susy-SUGRA}
Supergravity (SUGRA) \cite{sugra} is the theory that results from imposing 
local supersymmetry invariance. Recall that once we promote a global gauge 
symmetry (with bosonic generators, satisfying commutation relations) to a local
one, we have to introduce a bosonic field with predetermined 
gauge-transformation properties. Similarly, by promoting supersymmetry (which 
has \emph{fermionic} generators, satisfying \emph{anti}commutation relations) 
to a local symmetry, we have to introduce a \emph{fermionic} field $\Psi_\mu$ 
with spin-$3/2$. This is the gravitino, the superpartner of the spin-2 
graviton. The resulting SUGRA Lagrangian is non-renormalisable; there is as yet
no renormalisable quantum field theory of gravity. However, the 
non-renormalisable operators are suppressed by inverse powers of the Planck
mass $M_\text{Pl}={\cal O}(10^{19}\,\mbox{GeV})$, so that their effects at low 
energies are small (see section \ref{s:efts-efts}).

The spontaneous breakdown of supersymmetry occurs in a hidden sector where the 
auxiliary component of some superfield gets a VEV. According to Goldstone's 
theorem, spontaneously breaking a global symmetry yields a massless 
particle with the same quantum numbers as the broken symmetry generator.
Since the broken generator $Q$ is fermionic, the massless particle is a 
massless neutral Weyl fermion, called the \emph{goldstino}. The goldstino then 
becomes the longitudinal component of the gravitino, which becomes 
massive.\footnote{Because of the similarities with the Higgs mechanism, where 
the electroweak gauge bosons `eat' the Goldstone bosons and become massive, 
this mechanism is called the \emph{super-Higgs} mechanism.}
It turns out that when we consider the effects of the supersymmetry-breaking 
VEV, the gravitino mass $m_{3/2}$ sets the scale of all the soft terms.
Moreover, the scalar masses are universal at the scale where supersymmetry 
becomes broken.

\paragraph{Minimal supergravity.}
The most widely used model of supersymmetry breaking is minimal supergravity 
(mSUGRA) \cite{msugra1,msugra2}. Despite the name, mSUGRA is not a supergravity
model, but rather the low-energy EFT resulting from a minimal locally 
supersymmetric model. In the underlying model, one uses the simplest possible 
Ansatz for the scalar potentials. This leads to universal soft 
supersymmetry-breaking parameters in the scalar sector. Gauge-coupling 
unification in the MSSM suggests an additional simple Ansatz for the gauge 
kinetic function, which leads to universal gaugino masses. The model is then described
by four parameters and a sign. This involves three parameters at the GUT scale
$M_\text{GUT}=2\cdot10^{16}$ GeV: a universal scalar mass $m_0$, a universal gaugino mass 
$M_{1/2}$ and a universal proportionality factor $A_0$ between the trilinear 
couplings and the corresponding Yukawa couplings. In addition it involves the ratio $\tan\beta$ of the
two non-zero Higgs VEVs and the sign of the supersymmetric parameter 
$\mu$, which are both usually determined at the weak scale.
At the GUT scale, the soft terms relevant to our study are therefore 
given by
\begin{IEEEeqnarray}{rCl}\label{eq:msugraboundaryconditions}
m_i^2 &=& m_0^2, \IEEEyessubnumber\\
M_a &=& M_{1/2} \qquad (a=1,2,3), \IEEEyessubnumber
\end{IEEEeqnarray}
where $m_i^2$ are the scalar squared masses.
From the supergravity point of view, the parameters $m_0,M_{1/2},A_0$ 
depend on the hidden-sector fields and are all proportional to $m_{3/2}$ 
(for example, one has the relation $m_0=m_{3/2}$). However, from the 
perspective of the low-energy EFT that we call mSUGRA, they are simply regarded
as model parameters. The MSSM is assumed to be valid up to the GUT scale, where
the relations \eqref{eq:msugraboundaryconditions} serve as RG boundary 
conditions. In addition, the soft Higgs mixing term $B=b/\mu$ has the GUT-scale
value $B_0=A_0-m_{3/2}$.

As an aside, there is a model similar to mSUGRA: it is called constrained MSSM 
(CMSSM, see e.g.\ \cite{CMSSM}). It has the same boundary conditions as mSUGRA 
and these two models are often confused in the literature. However, mSUGRA 
arises from a supergravity model whereas the CMSSM does not: the boundary 
conditions \eqref{eq:msugraboundaryconditions} are simply postulated.
Also, in the CMSSM there is no relation between the model parameters and 
$m_{3/2}$, and the relation $B_0=A_0-m_{3/2}$ does not hold either.

Supergravity models are attractive since they provide a natural framework for 
supersymmetry breaking: a locally supersymmetric Lagrangian automatically 
contains terms that can mediate supersymmetry breaking.
In addition, mSUGRA has great predictive power since it has only four free
parameters plus a sign. However, these models must necessarily appeal to Planck-scale 
physics, which is still poorly understood. Furthermore, though gravity is 
flavour blind, the supergravity invariance of the Lagrangian cannot prevent the
occurrence of (Planck-scale suppressed) flavour-mixing operators that 
correspond to tree-level interactions between hidden-sector fields and 
visible-sector fields. In order to suppress sparticle-induced FCNC processes, 
one must resort to additional generation symmetries.

\subsection{Anomaly-mediated supersymmetry breaking}\label{s:susy-AMSB}
In some models of supergravity, the visible and hidden sectors are physically 
separated by extra dimensions \cite{AMSB1,AMSB2}. In these `braneworld' 
scenarios, often inspired by string theory, our four-dimensional world is 
embedded in a higher-dimensional bulk that has additional spatial dimensions, 
which are curled up.

The general idea is that the MSSM fields and the hidden-sector fields are 
confined to parallel, distinct three-branes (space-like hypersurfaces), 
separated by a distance $r$. Only the gravity supermultiplet (and possibly new 
heavy fields) resides in the bulk. In this scenario every flavour-violating 
term that plagues supergravity, caused by tree-level couplings with a bulk 
field of mass $m$, is suppressed by a factor $e^{-mr}$.
Provided that $r$ is large enough, the flavour-violating effects are 
exponentially suppressed without requiring any fine-tuning. This class of 
models is called Anomaly-Mediated Supersymmetry Breaking (AMSB), because the 
size of the soft supersymmetry-breaking terms is determined by the loop-induced
superconformal (Weyl) anomaly \cite{Weylanomaly}. Local superconformal 
invariance is a rescaling symmetry that is violated at the quantum level.

Anomaly-mediated terms are always present in supergravity, but they are 
loop-sup\-pressed with respect to the gravitino mass and hence result in
subleading-order contributions to the soft masses. AMSB is the scenario where 
there are no supergravity contributions at tree level, so that the 
anomaly-mediated terms become the dominant ones. At the scale 
$M_{_\text{AMSB}}$ where supersymmetry breaking occurs, the soft 
terms relevant to our study have the following values (using the usual pMSSM
assumptions):
\begin{IEEEeqnarray}{rCl}\label{eq:AMSBboundaryconditions} 
M_a &=& \frac{b_a}{16\pi^2}\,g_a^2(M_{_\text{AMSB}})m_{3/2} \qquad (a=1,2,3),
        \IEEEyessubnumber\\
m_i^2 &=& \frac12 \dot{\gamma}_i\,m_{3/2}^2. 
          \IEEEyessubnumber\label{eq:AMSBsoftscalars}
\end{IEEEeqnarray}
Here, $m_i^2$ are again the scalar squared masses with $\gamma_i$ being the 
corresponding anomalous dimensions and $b_a = (\frac{33}{5},1,-3)$\, for 
\,$a=1,2,3$. The derivatives $\dot{\gamma}_i\equiv \d\gamma_i/\d t$ are 
explicitly given by
\begin{IEEEeqnarray}{rCl}
(16\pi^2)^2\dot{\gamma}_{H_u} &=& 6|y_t|^2B_t - 3g_2^4 - \frac{99}{25}g_1^4, 
                                  \IEEEyessubnumber\\
(16\pi^2)^2\dot{\gamma}_{H_d} &=& 6|y_b|^2B_b + 2|y_\tau|^2B_\tau - 3g_2^4 
                                  - \frac{99}{25}g_1^4, \IEEEyessubnumber\\
(16\pi^2)^2\dot{\gamma}_{\widetilde{Q}_i} &=& \delta_{i3}\left(2|y_t|^2B_t
        +2|y_b|^2B_b\right) + 16g_3^4 - 3g_2^4 - \frac{11}{25}g_1^4, 
        \IEEEyessubnumber\\
(16\pi^2)^2\dot{\gamma}_{\widetilde{\bar{u}}_i} &=& \delta_{i3}\cdot4|y_t|^2B_t
        + 16g_3^4 - \frac{176}{25}g_1^4, \IEEEyessubnumber\\
(16\pi^2)^2\dot{\gamma}_{\widetilde{\bar{d}}_i} &=& \delta_{i3}\cdot4|y_b|^2B_b
        + 16g_3^4 - \frac{44}{25}g_1^4, \IEEEyessubnumber\\
(16\pi^2)^2\dot{\gamma}_{\widetilde{L}_i} &=& \delta_{i3}\cdot2|y_\tau|^2B_\tau
        - 3g_2^4 - \frac{99}{25}g_1^4, \IEEEyessubnumber\\
(16\pi^2)^2\dot{\gamma}_{\widetilde{\bar{e}}_i} &=& 
        \delta_{i3}\cdot4|y_\tau|^2B_\tau-\frac{396}{25}g_1^4, 
        \IEEEyessubnumber
\end{IEEEeqnarray}
where we have defined the following quantities for convenience:
\begin{IEEEeqnarray}{rCl}
B_t &\equiv& 6|y_t|^2 + |y_b|^2 - \frac{16}{3}g_3^2 - 3g_2^2 
             - \frac{13}{15}g_1^2, \IEEEyessubnumber\\
B_b &\equiv& 6|y_b|^2 + |y_t|^2 + |y_\tau|^2 - \frac{16}{3}g_3^2 - 3g_2^2 
             - \frac{7}{15}g_1^2, \IEEEyessubnumber\\
B_\tau &\equiv& 4|y_\tau|^2 + 3|y_b|^2 - 3g_2^2 - \frac95g_1^2. 
                \IEEEyessubnumber
\end{IEEEeqnarray}

\paragraph{Minimal anomaly mediation.}
The advantage of AMSB over SUGRA is that it naturally conserves flavour.
However, pure anomaly mediation leads to tachyonic sleptons, i.e.\ their 
squared soft masses become negative. This would cause them to acquire non-zero 
VEVs and break the electromagnetic gauge symmetry. The minimal AMSB (mAMSB) 
model uses a phenomenological approach to tackle this problem: a universal, 
non-anomaly-mediated contribution $m_0^2$ is added to the soft squared scalar 
masses \eqref{eq:AMSBsoftscalars} at the scale $M_{_\text{AMSB}}$. The origin 
of these terms may be for example additional fields in the bulk, but in the 
mAMSB model $m_0$ is simply considered as a parameter of the model.

\subsection{General gauge mediation}\label{s:susy-GGM}
\begin{figure}[t]
{\setlength{\unitlength}{1mm}
    \begin{minipage}{\textwidth}
      \begin{center}
      \fmfcmd{
      vardef cross_bar (expr p, len, ang) =
       ((-len/2,0)--(len/2,0))
       rotated (ang + angle direction length(p)/2 of p)
       shifted point length(p)/2 of p
      enddef;
      style_def dashes_crossed expr p=
       save dpp;
       numeric dpp;
       dpp = ceiling (pixlen (p, 10) / dash_len) / length p;
       for k=0 upto dpp*length(p) - 1:
        cdraw point k/dpp of p ..
         point (k+.5)/dpp of p;
       endfor
       ccutdraw cross_bar (p, 3mm, 45);
       ccutdraw cross_bar (p, 3mm, -45);
      enddef;}
      \begin{fmfgraph*}(70,20)
        \fmfleft{i1,i2}
        \fmfright{o1,o2}
        \fmf{dashes}{i1,v1,v2,o1}
        \fmffreeze
        \fmflabel{$\tilde{\text{u}}$}{i1}
        \fmflabel{$\tilde{\text{u}}$}{o1}
        \fmf{phantom}{i2,w1,w2,o2}
        \fmf{wiggly,tension=.5,label=$B$,label.side=left}{v1,w1}
        \fmf{wiggly,tension=.5,label=$B$}{v2,w2}
        \fmffreeze
        \fmf{dashes_crossed,width=2,left=.5,label=$\langle F_X\rangle$}{w1,w2,w1}
        \fmfforce{(.39w,.65h)}{messenger}
          \fmfv{label=$\phi_M$,label.angle=0}{messenger}
      \end{fmfgraph*}
      \end{center}
    \end{minipage}}
  \caption{Contribution to the soft squared mass of the up squark in models of 
           Gauge-Mediated Supersymmetry Breaking (GMSB). The auxiliary 
           component of the spurion field $X$ obtains a supersymmetry-breaking 
           VEV $\langle F_X\rangle$. The up squark only couples indirectly to 
           this VEV: the scalar component $\phi_M$ of a messenger 
           supermultiplet couples at tree level to the spurion and through the 
           MSSM gauge fields (in this diagram the $B$ boson) with the MSSM. 
           When the messengers are integrated out from the action, this diagram
           contributes to the soft mass of the up squark.}\label{f:GMSB}
\end{figure}
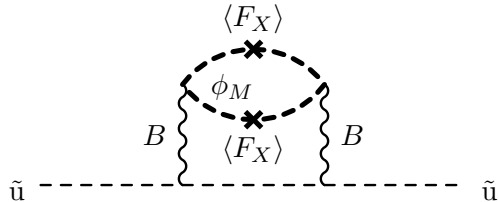

Several models of Gauge-Mediated Supersymmetry Breaking (GMSB) have been 
proposed in the literature (see \cite{GiudiceGMSB} for a review).
Many of these models include a field $X$, called the \emph{spurion}, that 
acquires a supersymmetry-breaking VEV, and a set of weakly coupled fields that 
are charged under the MSSM. The latter are called \emph{messenger} fields since
they communicate supersymmetry breaking to the MSSM fields: they interact at 
tree level with the spurion and through the MSSM gauge fields with the MSSM 
(see figure \ref{f:GMSB}).

Recently, the framework of General Gauge Mediation (GGM) \cite{GGM1,GGM2} has 
been proposed to unify all earlier descriptions of GMSB. It describes the 
effects of an arbitrary hidden sector on the MSSM. It starts from the following
definition of gauge mediation: \emph{in the limit of vanishing gauge couplings,
the theory decouples into the MSSM and a separate, supersymmetry-breaking 
hidden sector}. For example, the setup described above fits into this 
definition by taking the messenger and spurion fields as the hidden sector.

In the GGM framework, all MSSM soft terms can be described in terms of a small 
number of correlation functions involving hidden-sector currents. Essentially, 
the GGM framework parametrises the effects of the hidden sector on the MSSM.
By constructing the effective Lagrangian, the following soft-mass formulae are 
found:\footnote{For future convenience, a factor $M$ (the messenger scale) has 
been absorbed into the definition of the~$B_a$, cf.\ \cite{Carena2}.}
\begin{IEEEeqnarray}{rCl}
M_a &=& g_a^2 B_a \qquad (a=1,2,3), \IEEEyessubnumber\\
m_{i}^2 &=& g_1^2 Y_i\zeta + \sum_{a=1}^3 g_a^4 C_a(i) A_a, \IEEEyessubnumber\label{eq:GGMsfermionmasses}
\end{IEEEeqnarray}
with
\begin{IEEEeqnarray}{rCl}
C_1(i) &=& \frac35Y_i^2, \IEEEyessubnumber\\[2mm]
C_2(i) &=& \begin{cases} 
           \frac34 & \text{ for $\Phi_i=\widetilde{Q},\widetilde{L},H_u,H_d$,}\\
                 0 & \text{ for $\Phi_i=\widetilde{u}_R,\widetilde{d}_R,
                            \widetilde{e}_R$,} 
           \end{cases} \IEEEyessubnumber\\
C_3(i) &=& \begin{cases} 
           \frac43 & \text{ for $\Phi_i=\widetilde{Q},\widetilde{u}_R,
                            \widetilde{d}_R$,} \\ 
           0 & \text{ for $\Phi_i=\widetilde{L},\widetilde{e}_R,H_u,H_d$.} 
           \end{cases} \IEEEyessubnumber
\end{IEEEeqnarray}
Here $B_a,\zeta$ and $A_a$ are expressions involving the hidden-sector current 
correlation functions; $Y_i$ is the hypercharge of the scalar field $\Phi_i$ 
and $C_a(i)$ is the quadratic Casimir of the representation of 
$\Phi_i$ under the gauge group labeled by $a$. Usually a $\mathbf{Z}_2$ 
symmetry of the hidden sector is assumed in order to forbid the term containing
$\zeta$, since it would lead to tachyonic sleptons. The above conditions are 
the matching conditions at the messenger scale $M$ where we integrate out the 
hidden sector. The seven numbers $\zeta,A_a,B_a$ contain information on the 
hidden sector, but are regarded as parameters of the low-energy EFT that we 
call the MSSM.

The GGM framework does not allow for additional interactions that could 
generate $\mu$ and $b$ radiatively; that would require interactions between the
MSSM and the hidden sector that remain in the limit of vanishing gauge 
couplings. The framework would have to be extended to allow for such couplings.
To parametrise the effects of such an extension, additional contributions 
$\delta_u,\delta_d$ to $m_{H_u}^2,m_{H_d}^2$ are often added.

\paragraph{Minimal gauge mediation.}
Minimal gauge mediation (MGM) is a GGM model that is restricted to a subset of 
the GGM parameter space, defined by the constraints $A_a=A$, $B_a=B$ and
$A=2B^2$. The term corresponding to $\zeta$ is taken to be zero. Additional 
non-gauge contributions $\delta_u$, $\delta_d$ are added to the soft Higgs 
masses. Then the expressions for the soft masses become \vspace*{-2mm}
\begin{IEEEeqnarray}{rCl}\label{eq:MGMspectrum}
M_a &=& g_a^2B_a \qquad (a=1,2,3), \IEEEyessubnumber \label{eq:MGMgauginos}
        \\[2mm]
m_i^2 &=& 2B^2\sum_{a=1}^3 g_a^4C_a(i), 
          \IEEEyessubnumber \label{eq:MGMsfermions}\\
\mhu &=& 2B^2\sum_{a=1}^3 g_a^4C_a(H_u) + \delta_u, \IEEEyessubnumber \\
\mhd &=& 2B^2\sum_{a=1}^3 g_a^4C_a(H_d) + \delta_d, \IEEEyessubnumber
\end{IEEEeqnarray}
where this time $m_i^2$ denote only the squared masses of the squarks and 
sleptons.

\subsection{Mirage mediation}\label{s:susy-MM}
Rather than restricting oneself to one of the three known mechanisms for 
radiative supersymmetry breaking (gravity, anomaly or gauge mediation), one 
could solve the problems of particular models by choosing two (or more) 
mechanisms and combining the best of both worlds. For example, one might tackle
the tachyonic slepton problem of anomaly mediation by combining it with gauge 
mediation (see e.g.\ \cite{gauge+anomalymediation}).

Mirage mediation \cite{mirage} is one such scenario in which gravity-mediated 
and anomaly-mediated soft terms have comparable contributions. In this 
scenario, the gravity-mediated terms are suppressed by a relative factor 
$\log\left(M_\text{pl}/m_{3/2}\right)$, which is numerically of the order of a 
loop factor. This results in mirage unification: the gaugino and scalar masses 
unify at a scale far below the scale where the soft masses are generated.
This \emph{mirage messenger scale} does not correspond to any physical 
threshold, hence the name.

This class of phenomenological models are based on a class of string models 
with stabilised moduli, called the KKLT construction. It solves the tachyonic 
slepton problem that arises in pure anomaly mediation and has reduced 
low-energy fine-tuning \cite{miragefinetuning}.

\subsection{One-loop RGIs for the MSSM}\label{s:susy-mssmrgis}
It should be noted that if we have a set of RGIs, then any function of those 
RGIs will also be RG invariant. Therefore, in order to find all RGIs, one 
should look for a maximal set of \emph{independent} RGIs, i.e.\ invariants that
cannot be expressed in terms of each other.\footnote{It is tempting to call 
this a `basis of RGIs', as in \cite{Carena1}. Note however that it is not the 
same as a basis of a vector space. One should keep in mind that once we have 
found such a set, we are not restricted to making linear combinations of them, 
but can also take products, quotients and so on.} Recently a complete list of 
independent one-loop RGIs for the MSSM was derived in \cite{Demir,Carena1}: 
\begin{IEEEeqnarray}{rCl}\label{eq:mssmrgis}
D_{B_{13}} &=& 2\left(\msqa-\msqc\right) - \msua + \msuc - \msda + \msdc, \IEEEyessubnumber\label{eq:db13}\\
D_{L_{13}} &=& 2\left(\msla-\mslc\right) - \msea + \msec, \IEEEyessubnumber\label{eq:dl13}\\
D_{\chi_1} &=& 3\left(3\msda-2\left(\msqa-\msla\right)-\msua\right) - \msea, \IEEEyessubnumber\label{eq:dchi1}\\
D_{Y_{13H}} &=& \msqa-2\msua+\msda-\msla+\msea \nonumber\\
&&-\frac{10}{13}\left(\msqc-2\msuc+\msdc-\mslc+\msec+\mhu-\mhd\right), \IEEEyessubnumber\label{eq:dy13h}\\
D_Z &=& 3\left(\msdc-\msda\right) + 2\left(\mslc-\mhd\right), \IEEEyessubnumber\label{eq:dz}\\
I_{Y_\alpha} &=& \frac{1}{g_1^2}\left(\mhu-\mhd + \sum_\text{gen}\left(\msq-2\msu+\msd-\msl+\mse\right)\right), \IEEEyessubnumber\label{eq:iyalpha}\\
I_{B_a} &=& \frac{M_a}{g_a^2} \qquad (a=1,2,3),\IEEEyessubnumber\label{eq:iba}\\
I_{M_1} &=& M_1^2 - \frac{33}{8}\left(\msda-\msua-\msea\right), \IEEEyessubnumber\label{eq:im1}\\
I_{M_2} &=& M_2^2 + \frac{1}{24}\left(9\left(\msda-\msua\right) + 16\msla-\msea\right), \IEEEyessubnumber\label{eq:im2}\\
I_{M_3} &=& M_3^2 - \frac{3}{16}\left(5\msda+\msua-\msea\right), \IEEEyessubnumber\label{eq:im3}\\
I_{g_2} &=& g_1^{-2}-\frac{33}{5}g_2^{-2}, \IEEEyessubnumber\label{eq:ig2}\\
I_{g_3} &=& g_1^{-2}+\frac{11}{5}g_3^{-2}, \IEEEyessubnumber\label{eq:ig3}\\
I_2 &=& \mu\left( \frac{g_2^9g_3^{256/3}}{y_t^{27}y_b^{21}y_\tau^{10}g_1^{73/33}} \right)^{1/61}, \IEEEyessubnumber\label{eq:i2}\\
I_4 &=& \frac{b}{\mu} - \frac{27}{61}A_t - \frac{21}{61}A_b - \frac{10}{61}A_\tau - \frac{256}{183}M_3 - \frac{9}{61}M_2 + \frac{73}{2013}M_1. \IEEEyessubnumber\label{eq:i4}
\end{IEEEeqnarray}
The sum in~\eqref{eq:iyalpha} runs over the three sfermion generations.
The notation has been taken over from \cite{Carena1} (equations~\eqref{eq:db13}--\eqref{eq:ig3}) and \cite{Demir} (equations~\eqref{eq:i2}--\eqref{eq:i4}).
These one-loop RGIs have been derived under the pMSSM assumptions, using the relevant $\beta$-functions 
listed in appendix~\ref{a:mssmrges}. An alternative, systematic way of deriving
this complete set can be found in appendix~\ref{a:mssmrgis}.

Any other (one-loop) RGI we can think of can be written in terms of those in~\eqref{eq:mssmrgis}. 
For example, the RGI in the example from 
section~\ref{s:efts-rgis} can be written as
\begin{align}
I_g &= 4g_1^{-2} - \frac{48}{5}g_2^{-2} + \frac{28}{5}g_3^{-3} \nonumber\\
    &= \frac{16}{11}I_{g_2} + \frac{28}{11}I_{g_3}.
\end{align}
The two RGIs in~\eqref{eq:i2}--\eqref{eq:i4} will not be relevant to our 
analysis. We will explain why in the next subsection.

\subsection{RGIs in the literature}\label{s:susy-rgiliterature}
As we have seen, RGIs provide a new tool to test predictions about high-scale 
physics, such as gauge-coupling unification. The trick is to find sum rules for
high-scale physics that can be written in terms of RGIs. In the literature, 
several such sum rules can be found.

Consider for example minimal gauge mediation, which has been studied in the 
context of RGIs in \cite{Carena2}. If one inserts the spectrum 
\eqref{eq:MGMspectrum} at the messenger scale into the RGI expressions 
from~\eqref{eq:db13}--\eqref{eq:ig3}, one immediately finds that 
$D_{B_{13}} = D_{L_{13}} = D_{\chi_1} = 0$. In terms of the model parameters of
MGM, the non-vanishing RGIs are
\begin{IEEEeqnarray}{rCl}
D_{Y_{13H}} &=& -\frac{10}{13}\left(\delta_u-\delta_d\right), 
                \IEEEyessubnumber\label{eq:mgmrgi1}\\
D_Z &=& -2\delta_d, \IEEEyessubnumber\label{eq:mgmrgi2}\\
I_{Y_\alpha} &=& g_1^{-2}(M)\left(\delta_u-\delta_d\right), 
                 \IEEEyessubnumber\label{eq:mgmrgi3}\\
I_{B_a} &=& B \qquad (a=1,2,3), \IEEEyessubnumber\label{eq:mgmrgi4}\\
I_{M_1} &=& \frac{38}{5}g_1^4(M)B^2, \IEEEyessubnumber\label{eq:mgmrgi5}\\
I_{M_2} &=& 2g_2^4(M)B^2, \IEEEyessubnumber\label{eq:mgmrgi6}\\
I_{M_3} &=& -2g_3^4(M)B^2, \IEEEyessubnumber\label{eq:mgmrgi7}\\
I_{g_2} &=& g_1^{-2}(M) -\frac{33}{5}g_2^{-2}(M), 
            \IEEEyessubnumber\label{eq:mgmrgi8}\\
I_{g_3} &=& g_1^{-2}(M) + \frac{11}{5}g_3^{-2}(M) ,
            \IEEEyessubnumber\label{eq:mgmrgi9}
\end{IEEEeqnarray}
where $M$ is the messenger scale. This gives us eleven equations in terms of 
six unknowns ($\delta_u$, $\delta_d$, $B$, $g_1(M)$, $g_2(M)$, $g_3(M)$).
We can trade each unknown for an equation, i.e.\ for  each parameter we use one 
of the above equations to express it in terms of RGIs only. Since we have more 
independent equations than unknowns, we can substitute the resulting six 
expressions into the remaining five equations to obtain five sum rules in terms
of RGIs only. For example, using equation \eqref{eq:mgmrgi4} we can eliminate 
the model parameter $B$ from the remaining equations. Then equations 
\eqref{eq:mgmrgi5}-\eqref{eq:mgmrgi7} can be used to eliminate the gauge 
couplings at the messenger scale. We can get the value of $\delta_d$ from 
equation \eqref{eq:mgmrgi2}, and then \eqref{eq:mgmrgi1} gives the value of 
$\delta_u$. After substituting the resulting six expressions into the five 
remaining equations, we are left with the following sum rules:
\begin{align}
0 &= I_{Y_\alpha}+\frac{13}{10}D_{Y_{13H}}I_{B_1}\sqrt{\frac{38}{5I_{M_1}}}\,,
     \tag{\ref{eq:mgmrgi3}}\\
0 &= I_{B_1} - I_{B_2}, \tag{\ref{eq:mgmrgi4}}\\[3mm]
0 &= I_{B_1} - I_{B_3}, \tag{\ref{eq:mgmrgi4}}\\
0 &= I_{B_1}\sqrt{\frac{38}{5I_{M_1}}} 
     - \frac{33}{5}I_{B_1}\sqrt{\frac{2}{I_{M_2}}} - I_{g_2}, 
     \tag{\ref{eq:mgmrgi8}}\\
0 &= I_{B_1}\sqrt{\frac{38}{5I_{M_1}}} 
     + \frac{11}{5}I_{B_1}\sqrt{\frac{-2}{I_{M_3}}} - I_{g_3}, 
     \tag{\ref{eq:mgmrgi9}}
\end{align}
To summarise, we have chosen a specific supersymmetry-breaking model and 
expressed the RGIs in terms of model parameters. Since we ended up with more 
equations than unknowns, we could construct eight sum rules in terms of RGIs 
only: three from vanishing RGIs and five by eliminating the model parameters.
If any of these sum rules are violated, MGM is not consistent with experimental
data. This test can be performed at any energy scale where the MSSM is valid,
which implies that we don't have to know the messenger scale $M$ in order to 
rule out the MGM model.

Now we can see why the two RGIs in~\eqref{eq:i2}--\eqref{eq:i4} are not useful. 
Suppose we wish to test a specific breaking model. Let us denote the values of 
$B=b/\mu$ and $\mu$ at the new-physics threshold in this model as 
$B_\text{thr}$ and $\mu_\text{thr}$ respectively. Now we apply the above 
procedure to this model: we express $B_\text{thr}$ and $\mu_\text{thr}$ in 
terms of RGIs and the other couplings at the high scale; then we can insert 
these expressions into the remaining equations. But since $B$ and $\mu$ both 
appear in only one independent RGI, there are no equations to insert these 
expressions into! In the above example, it was possible to combine all RGIs 
into sum rules because each running parameter appeared in more than one RGI.
Since $B$ and $\mu$ do not, their corresponding RGIs become useless to our 
analysis. Hence we will have to restrict ourselves to the RGIs that do not 
contain these running parameters. As will be shown in 
appendix~\ref{a:mssmrgis}, it is not possible to construct RGIs that contain 
the Yukawa and soft trilinear couplings without using $B$ and $\mu$. That is 
why we will only use RGIs constructed out of soft masses and/or gauge 
couplings, i.e.~the RGIs listed in~\eqref{eq:db13}--\eqref{eq:ig3}.

\subsubsection{Using RGIs effectively}\label{s:susy-rgieffectively}
In studies of RGIs such as \cite{Carena1,Carena2,Jaeckel}, a certain breaking 
mechanism is usually presupposed. Then one constructs sum rules that are 
tailor-made for that breaking mechanism. For example, the sum rules constructed
above all provide a test for consistency of MGM with experimental data.
However, some of these sum rules will also hold for other breaking mechanisms.
It is not always clear to what extent the validity of the sum rules depends on 
the unique features of the breaking mechanism under study. For example, in 
minimal gauge mediation the quantity $M_a/g_a^2$ unifies at the messenger 
scale; this follows directly from the matching condition 
\eqref{eq:MGMgauginos}. However, in mSUGRA this quantity also unifies, but for 
a different reason: it is the consequence of the assumption of gauge-coupling 
unification and gaugino-mass unification at the same energy scale!
Hence, the sum rules that test this unification property cannot be used to 
confirm that either of these specific models corresponds to reality. They can 
only provide consistency checks that should be satisfied if any of these models
are realised in Nature.

Therefore, we will look for RG-invariant sum rules using a different, more
model-independent approach. We will not presume any spectrum specific to a 
certain breaking mechanism. Instead, we will search for sum rules that test 
properties that are common in supersymmetry-breaking models 
(e.g.\ $M_a/g_a^2$ unification). Then any breaking model, be it an existing one
such as those described in sections \ref{s:susy-SUGRA}--\ref{s:susy-MM} or a 
new one contrived in the future, can be tested directly if it predicts any of 
these properties. For example, if the sum rules for $M_a/g_a^2$ unification are
not satisfied by experimental data, then models that predict this property 
(mSUGRA and MGM, but not necessarily GGM) are falsified. But also anyone who 
would concoct a new model that has this property, would have to go back to the 
drawing board at once. In the next section, we will look for common properties 
to test and find sum rules for them.

\end{fmffile}

\section{Results}\label{s:results}

Supersymmetry-breaking models predict relations between the running parameters 
as a result of matching conditions at the new-physics threshold.
These relations mostly involve the unification of certain parameters.
Therefore we will construct sum rules for the following scenarios.

\paragraph{Scenario~1: gauge-coupling unification.}
As can be seen from figure~\ref{f:gaugeunification}, the MSSM may be consistent
with gauge-coupling unification, depending on the values of the sparticle 
thresholds. The hypothesis that the gauge couplings unify is often made in 
supersymmetry-breaking models, for example in mSUGRA. Therefore it will be 
important to determine whether gauge-coupling unification occurs in Nature.
We will call the special case where $g_1=g_2=g_3=g$ scenario~1.

\paragraph{Scenario~2: gaugino-mass unification.}
In mSUGRA, the gaugino masses are assumed to unify. Since this model is widely 
used, it is useful to check whether gaugino-mass unification occurs in 
Nature. We will call the case where $M_1=M_2=M_3=M_{1/2}\,$~scenario~2.

\paragraph{Scenario~3: Unification of $\mathbf{M_a/g_a^2}$.}
As we mentioned in section~\ref{s:susy}, the quantities $M_1/g_1^2,\,M_2/g_2^2$
and $M_3/g_3^2$  may unify for different reasons. It could be the consequence 
of gaugino-mass and gauge-coupling unification at the same scale (as in mSUGRA)
or it may be the result of the gaugino-mass matching conditions (as in MGM). 
Therefore we will also test this property. We will call the case where 
$M_1/g_1^2=M_2/g_2^2=M_3/g_3^2=C\,$ scenario~3.

\paragraph{Scenario~4: flavour-universality of high-scale sfermion masses.}
In many theories, the sfermion masses are assumed to be flavour-universal, 
i.e.\ the first (and therefore also the second) and third generation masses are
equal: \mbox{$\msqa=\msqc\equiv\msq$\,,} \mbox{$\msua=\msuc\equiv\msu$\,,} 
\mbox{$\msda=\msdc\equiv\msd$\,,} \mbox{$\msla=\mslc\equiv\msl$} and 
\mbox{$\msea=\msec\equiv\mse$\,.} We will call this scenario~4.
This hypothesis is motivated by the need to suppress FCNC amplitudes.
Flavour-universality may be a consequence of flavour symmetries (as postulated 
in mSUGRA) or of the flavour-blindness of the interactions that mediate 
supersymmetry breaking (as in GGM). Since this property occurs in many models, 
we will test flavour-universality of the soft masses.

\paragraph{Scenario 5: scalar-mass unification.}
Unification of the soft scalar masses, which occurs in mSUGRA, is very 
predictive: many matching conditions depend on a single parameter $m_0$, which 
allows us to construct multiple sum rules. Sometimes non-universality of the 
soft Higgs masses is assumed, because suppression of FCNC amplitudes does not 
require them to be universal with the squark and slepton masses. Therefore we 
will distinguish between two cases. We will refer to the case where $\mhu$ and 
$\mhd$ have additional non-universal contributions $\delta_u$ and $\delta_d$ 
respectively as scenario~5a. The special case of universal scalar masses, 
i.e.\ $\delta_u=\delta_d=0$, will be denoted as scenario~5b.

\paragraph{Mixed scenarios: multiple unifications at one scale.}
It is possible that several of these unification properties will turn out to be
consistent with experimental data. In that case, we could test whether these 
unifications occur at the same energy scale. At first sight, it may seem 
strange to consider the possibility of two kinds of unifications at different 
energy scales. In supersymmetry-breaking models such unifications usually occur
at a threshold where new physics enters the theory. Thus even if the MSSM were 
consistent with two kinds of unifications, the RG trajectories of the running 
parameters could be deflected from the MSSM trajectories after the first 
threshold, spoiling the second unification. However, recall that in mirage 
mediation (see section~\ref{s:susy-MM}) the scale where the soft masses unify 
is lower than the scale at which the soft masses are generated. Thus it is 
possible that the unification scale does not correspond to any physical 
threshold. Therefore we will separately check whether multiple unifications 
occur at the same scale.

\subsection{Sum rules}\label{s:results-sumrules}
In this section we construct sum rules that test the scenarios described above. 

\paragraph{Scenario 1.}
From equations~\eqref{eq:db13}--\eqref{eq:ig3} and table~\ref{t:rgivalues} we can see that 
we have 14 equations with 16 unknowns, so at first sight we expect to find no 
sum rules. However, equations \eqref{eq:ig2} and \eqref{eq:ig3} form a subset 
of two equations with only one unknown $g$. Hence we can make one sum rule:
\begin{equation}
I_{g_2}+\frac74I_{g_3}=0. \label{eq:sumrule1}
\end{equation}

\paragraph{Scenario 2.}
In this scenario we also have 14 equations with 16 unknowns. However, equations
\eqref{eq:iba}, \eqref{eq:ig2} and \eqref{eq:ig3} form a subset of five 
equations with four unknowns. This allows us to construct one sum rule:
\begin{equation}
\left(I_{B_1}-\frac{33}{5}I_{B_2}\right)I_{g_3} 
= \left(I_{B_1}+\frac{11}{5}I_{B_3}\right)I_{g_2}. \label{eq:sumrule2}
\end{equation}

\paragraph{Scenario 3.}
Again we have 14 equations with 16 unknowns. This time equations \eqref{eq:iba}
form a subset of three equations with one unknown. This yields two sum rules:
\begin{align}
I_{B_1} &= I_{B_2}, \label{eq:sumrule3.1}\\
I_{B_1} &= I_{B_3}. \label{eq:sumrule3.2}
\end{align}

\paragraph{Scenario 4.}
We can see directly from table~\ref{t:rgivalues} that this scenario yields two 
sum rules in the form of vanishing RGIs:
\begin{align}
D_{B_{13}} &= 0, \label{eq:sumrule4.1}\\
D_{L_{13}} &= 0. \label{eq:sumrule4.2}
\end{align}
The remaining 12 equations contain 13 unknowns and no subset of them contains 
less unknowns than equations. Hence, we cannot construct any other sum rules.

\paragraph{Scenario 5a.}
In this scenario we also have two vanishing RGIs, because unified scalar masses
imply flavour-universal scalar masses. The remaining RGIs yield twelve 
equations with nine unknowns. Hence, we can construct three sum rules that are 
specific to this scenario:
\begin{align}
I_{g_2} &= I_{B_1}\left(I_{M_1}-\frac{33}{40}D_{\chi_1}\right)^{-1/2} 
           - \frac{33}{5}I_{B_2}\left(I_{M_2}-\frac18D_{\chi_1}\right)^{-1/2}, 
           \label{eq:sumrule5a.1}\\
I_{g_3} &= I_{B_1}\left(I_{M_1}-\frac{33}{40}D_{\chi_1}\right)^{-1/2} 
           + \frac{11}{5}I_{B_3}\left(I_{M_3}+\frac{3}{16}D_{\chi_1}
                                \right)^{-1/2}, \label{eq:sumrule5a.2}\\[1mm]
0 &= I_{Y_\alpha}\sqrt{I_{M_1}-\frac{33}{40}D_{\chi_1}}
     + \frac{13}{10}I_{B_1}D_{Y_{13H}}. \label{eq:sumrule5a.3}
\end{align}
Furthermore, non-universality of the Higgs masses can be tested directly 
because we can extract $\delta_u$ and $\delta_d$ from the RGIs:
\begin{align}
\delta_d &= -\frac12D_Z\neq0, \label{eq:sumrule5a.4}\\
\delta_u &= -\frac{13}{10}D_{Y_{13H}}-\frac12D_Z \neq0. \label{eq:sumrule5a.5}
\end{align}

\begin{sidewaystable}[H]
{\renewcommand{\arraystretch}{2}
\begin{tabular}{|c|c|c|c|c|c|c|}
\hline
\textbf{Invariant}	&	\textbf{Scenario 1}	&	\textbf{Scenario 2}	&	\textbf{Scenario 3}	&	\textbf{Scenario 4}	&	\textbf{Scenario 5a}	&	\textbf{Scenario 5b}	\\\hline\hline
$D_{B_{13}}$	&	$D_{B_{13}}$	&	$D_{B_{13}}$	&	$D_{B_{13}}$	&	{\boldmath $0$}	&	$0$	&	$0$	\\\hline
$D_{L_{13}}$	&	$D_{L_{13}}$	&	$D_{L_{13}}$	&	$D_{L_{13}}$	&	{\boldmath $0$}	&	$0$	&	$0$	\\\hline
$D_{\chi_1}$	&	$D_{\chi_1}$	&	$D_{\chi_1}$	&	$D_{\chi_1}$	&	$D_{\chi_1}\left(m_{\widetilde{f}_i}^2\rightarrow m_{\widetilde{f}}^2\right)$	&	{\boldmath $5m_0^2$}	&	{\boldmath $5m_0^2$}	\\\hline
$D_{Y_{13H}}$	&	$D_{Y_{13H}}$	&	$D_{Y_{13H}}$	&	$D_{Y_{13H}}$	&	$D_{Y_{13H}}\left(m_{\widetilde{f}_i}^2\rightarrow m_{\widetilde{f}}^2\right)$	&	{\boldmath $-\frac{10}{13}(\delta_u-\delta_d)$}	&	{\boldmath $0$}	\\\hline
$D_Z$	&	$D_Z$	&	$D_Z$	&	$D_Z$	&	$D_Z\left(m_{\widetilde{f}_i}^2\rightarrow m_{\widetilde{f}}^2\right)$	&	{\boldmath $-2\delta_d$}	&	{\boldmath $0$}	\\\hline
$I_{Y_\alpha}$	&	$I_{Y_\alpha}\left(g_1\rightarrow g\right)$	&	$I_{Y_\alpha}$	&	$I_{Y_\alpha}$	&	$I_{Y_\alpha}\left(m_{\widetilde{f}_i}^2\rightarrow m_{\widetilde{f}}^2\right)$	&	{\boldmath $\frac{\delta_u-\delta_d}{g_1^2}$}	&	{\boldmath $0$}	\\\hline
$I_{B_a}\quad a=1,2,3$	&	$I_{B_a}\left(g_a\rightarrow g\right)$	&	{\boldmath $\frac{M_{1/2}}{g_a^2}$}	&	{\boldmath $C$}	&	$I_{B_a}$	&	{\boldmath $I_{B_a}$}	&	{\boldmath $I_{B_a}$}	\\\hline
$I_{M_1}$	&	$I_{M_1}$	&	$I_{M_1}\left(M_1\rightarrow M_{1/2}\right)$	&	$I_{M_1}\left(M_1\rightarrow Cg_1^2\right)$	&	$I_{M_1}\left(m_{\widetilde{f}_i}^2\rightarrow m_{\widetilde{f}}^2\right)$	&	{\boldmath $M_1^2+\frac{33}{8}m_0^2$}	&	{\boldmath $M_1^2+\frac{33}{8}m_0^2$}	\\\hline
$I_{M_2}$	&	$I_{M_2}$	&	$I_{M_2}\left(M_2\rightarrow M_{1/2}\right)$	&	$I_{M_2}\left(M_2\rightarrow Cg_2^2\right)$	&	$I_{M_2}\left(m_{\widetilde{f}_i}^2\rightarrow m_{\widetilde{f}}^2\right)$	&	{\boldmath $M_2^2+\frac58m_0^2$}	&	{\boldmath $M_2^2+\frac58m_0^2$}	\\\hline
$I_{M_3}$	&	$I_{M_3}$	&	$I_{M_3}\left(M_3\rightarrow M_{1/2}\right)$	&	$I_{M_3}\left(M_3\rightarrow Cg_3^2\right)$	&	$I_{M_3}\left(m_{\widetilde{f}_i}^2\rightarrow m_{\widetilde{f}}^2\right)$	&	{\boldmath $M_3^2-\frac{15}{16}m_0^2$}	&	{\boldmath $M_3^2-\frac{15}{16}m_0^2$}	\\\hline
$I_{g_2}$	&	{\boldmath $-\frac{28}{5}g^{-2}$}	&	{\boldmath $I_{g_2}$}	&	$I_{g_2}$	&	$I_{g_2}$	&	{\boldmath $I_{g_2}$}	&	{\boldmath $I_{g_2}$}	\\\hline
$I_{g_3}$	&	{\boldmath $\frac{16}{5}g^{-2}$}	&	{\boldmath $I_{g_3}$}	&	$I_{g_3}$	&	$I_{g_3}$	&	{\boldmath $I_{g_3}$}	&	{\boldmath $I_{g_3}$}	\\\hline
\end{tabular}}
\caption{Values of the MSSM RGIs in the unification scenarios 1--5. For each 
         scenario, RGIs that contribute to sum rules that are specific to that 
         scenario are listed in boldface. If an RGI is simplified with respect 
         to its definition but is not used for a sum rule, its name with the 
         appropriate substitutions is listed. For the substitution 
         $\bigl(m_{\widetilde{f}_i}^2\rightarrow m_{\widetilde{f}}^2\bigr)$ it 
         is implied that $\widetilde{f}=\widetilde{Q}, \widetilde{\bar{u}}, 
         \widetilde{\bar{d}}, \widetilde{L}, \widetilde{\bar{e}}\,$ and 
         $\,i=1,2,3$. If an RGI does not simplify at all, only its name is 
         listed.}\label{t:rgivalues}
\end{sidewaystable}

\paragraph{Scenario 5b.}
We can see directly from table~\ref{t:rgivalues} that this scenario yields 
three new sum rules in the form of vanishing RGIs:
\begin{align}
D_{Y_{13H}} &= 0, \label{eq:sumrule5b.1}\\
D_Z &= 0, \label{eq:sumrule5b.2}\\
I_{Y_\alpha} &= 0. \label{eq:sumrule5b.3}
\end{align}
The remaining nine equations contain seven unknowns, which allows us to 
construct two more sum rules:
\begin{align}
I_{g_2} &= I_{B_1}\left(I_{M_1}-\frac{33}{40}D_{\chi_1}\right)^{-1/2} 
           - \frac{33}{5}I_{B_2}\left(I_{M_2}-\frac18D_{\chi_1}\right)^{-1/2}, 
           \label{eq:sumrule5b.4}\\
I_{g_3} &= I_{B_1}\left(I_{M_1}-\frac{33}{40}D_{\chi_1}\right)^{-1/2} 
           + \frac{11}{5}I_{B_3}\left(I_{M_3}+\frac{3}{16}D_{\chi_1}
                                \right)^{-1/2}. \label{eq:sumrule5b.5}
\end{align}
Note that these two sum rules also hold for scenario~5a.
This makes sense, because scenario~5b is a special case of scenario~5a, so the 
sum rules \eqref{eq:sumrule5a.1}-\eqref{eq:sumrule5a.3} will also hold for 
scenario~5b. However, sum rule \eqref{eq:sumrule5a.3} has become redundant 
because it is automatically satisfied if \eqref{eq:sumrule5b.1} and 
\eqref{eq:sumrule5b.3} hold.

\paragraph{Mixed scenario 123.}
Note that if any two of the scenarios~1, 2 and 3 hold, the hypothesis that the 
corresponding unifications occur at the same scale is equivalent to the 
hypothesis that the third scenario also holds. For example, if the gauge 
couplings and the gaugino masses unify, these unifications occur at the same 
scale if and only if $M_a/g_a^2$ unification occurs. Hence, for simultaneous 
unification at least the sum rules \eqref{eq:sumrule1}--\eqref{eq:sumrule3.2} 
should hold. Note that \eqref{eq:sumrule2} is automatically satisfied if the 
other three sum rules hold, so we have three independent sum rules for this 
mixed scenario.

From table~\ref{t:rgivalues} we conclude that if these three unifications occur
simultaneously we have 14 equations with 14 unknowns, which gives no sum rules 
at first sight. However, equations \eqref{eq:iba}, \eqref{eq:ig2} and 
\eqref{eq:ig3} constitute five equations with two unknowns, giving us three sum
rules. We have already found three, so there are no new sum rules.

\paragraph{Mixed scenario 15.}
If both scenario~1 and scenario~5a hold, at least the sum rules 
\eqref{eq:sumrule1} and \eqref{eq:sumrule4.1}--\eqref{eq:sumrule5a.3} hold, as 
well as the inequalities \eqref{eq:sumrule5a.4}--\eqref{eq:sumrule5a.5}.
From table~\ref{t:rgivalues} we conclude that if these unifications occur 
simultaneously we have two sum rules from vanishing RGIs as well as twelve 
equations with seven unknowns. Thus we can construct five additional sum rules.
We already found six of them, so there is one new sum rule:
\begin{equation}
I_{Y_\alpha}=\frac{13}{56}I_{g_2}D_{Y_{13H}}. \label{eq:sumrule1+5}
\end{equation}
If both scenario~1 and scenario~5b hold, we have 
$D_Z=I_{Y_\alpha}=D_{Y_{13H}}=0$. In that case this new sum rule becomes 
redundant.

\paragraph{Mixed scenario 25.}
If both scenario~2 and scenario~5a hold, at least the sum rules 
\eqref{eq:sumrule2} and \eqref{eq:sumrule4.1}--\eqref{eq:sumrule5a.3} hold, as 
well as the inequalities \eqref{eq:sumrule5a.4}--\eqref{eq:sumrule5a.5}.
From table~\ref{t:rgivalues} we conclude that if these unifications occur 
simultaneously we have two sum rules from vanishing RGIs as well as twelve 
equations with seven unknowns. Hence we can construct five additional sum 
rules. We already found six of them, so there is one new sum rule:
\begin{equation}
I_{M_1}-\frac{81}{25}I_{M_2}+\frac{56}{25}I_{M_3} = 0 \label{eq:sumrule2+5}
\end{equation}
Note that we get the same additional sum rule if we choose scenario~5b instead 
of 5a.

\paragraph{Mixed scenario 125.}
If all unifications occur simultaneously, then we immediately know that the 
following sum rules should hold (we assume scenario~5a for the moment):
\begin{itemize}
\item Sum rule \eqref{eq:sumrule1} for scenario~1.
\item Sum rules \eqref{eq:sumrule3.1}--\eqref{eq:sumrule3.2} for scenario~3.
\item Sum rule \eqref{eq:sumrule2} for scenario~2. However, as we have 
      mentioned before, this one is automatically satisfied if the sum rules 
      for scenarios~1 and 3 are satisfied. Hence this one is redundant.
\item Sum rules \eqref{eq:sumrule4.1}--\eqref{eq:sumrule4.2} for scenario~4, 
      which is implied by scenario~5.
\item Sum rules \eqref{eq:sumrule5a.1}--\eqref{eq:sumrule5a.3} for scenario~5, 
      as well as inequalities \eqref{eq:sumrule5a.4}--\eqref{eq:sumrule5a.5}.
\item Sum rule \eqref{eq:sumrule1+5} for the simultaneity of scenarios~1 and 5.
\item Sum rule \eqref{eq:sumrule2+5} for the simultaneity of scenarios~2 and 5.
      However, this one has become redundant: it can be retrieved by combining 
      equations \eqref{eq:sumrule3.1}--\eqref{eq:sumrule3.2}, 
      \eqref{eq:sumrule5a.1}--\eqref{eq:sumrule5a.3} and \eqref{eq:sumrule1+5}.
      We could have expected this: if we have consistency with scenarios~1, 2 
      and 5, and we have established simultaneity of both scenarios~1 and 2 
      and scenarios~1 and 5, then it follows automatically that we have 
      simultaneity of scenarios~2 and 5.
\end{itemize}
This adds up to nine independent sum rules. From table~\ref{t:rgivalues} we 
conclude that if all unifications occur simultaneously we have two sum rules 
from vanishing RGIs as well as twelve equations with five unknowns. Hence, we 
should get nine sum rules, so there are no new ones. Note that if we had taken 
scenario~5b instead of 5a, we would not get any additional new sum rules 
either. In that case the usual sum rules $D_Z=I_{Y_\alpha}=D_{Y_{13H}}=0$ would
hold and equation \eqref{eq:sumrule1+5} would become redundant.

\subsubsection{Sum rules summary}
\begin{figure}[p]
\begin{center}
  \includegraphics[width=\textwidth]{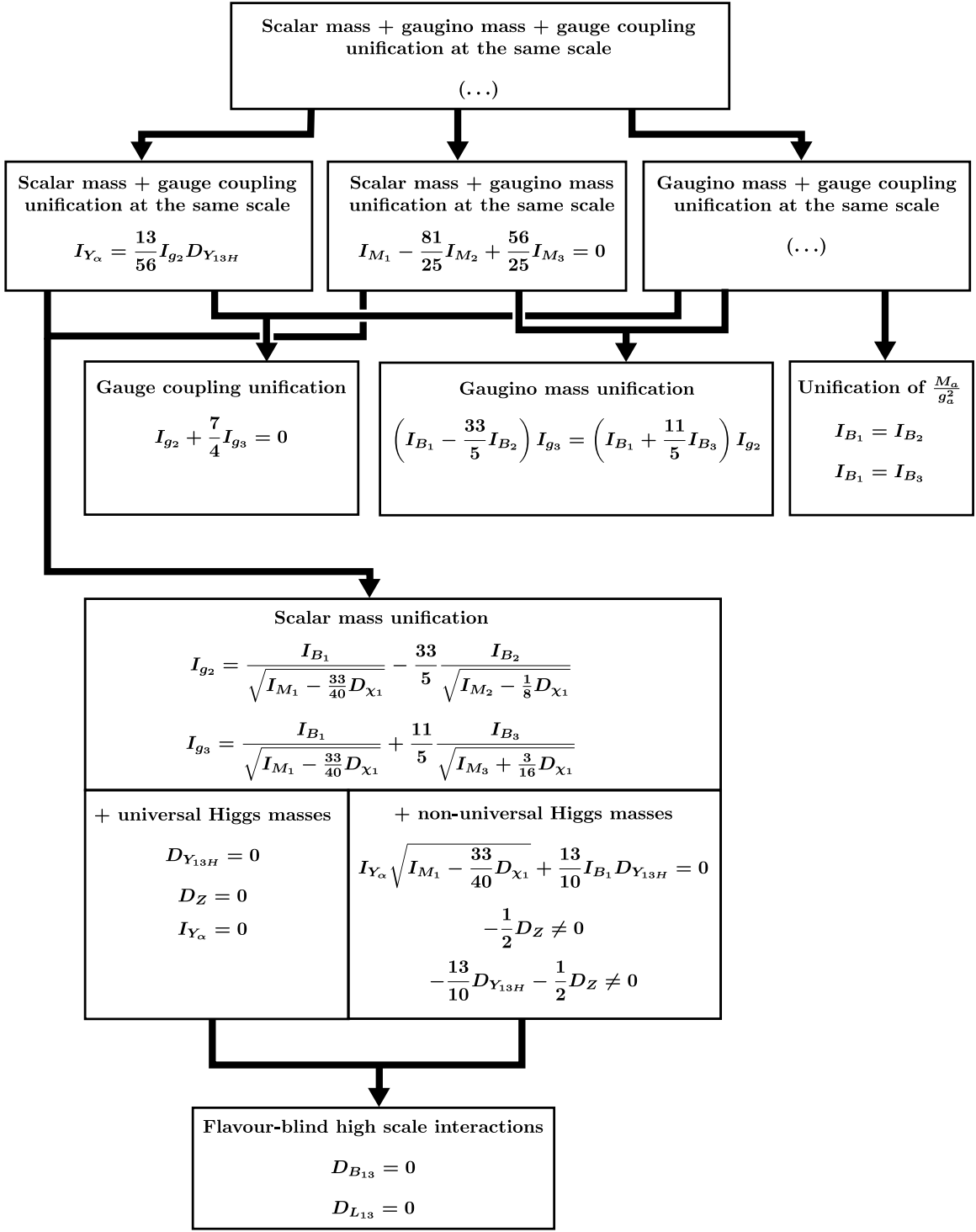}
\end{center}
\caption{Scheme for testing hypotheses about the spectrum at the new-physics 
         threshold. For a given scenario, the arrows point towards its 
         underlying hypotheses. To test a specific hypothesis, check whether 
         the corresponding sum rules are satisfied. Then follow the arrows 
         downwards all the way to the bottom and for each sum rule along the 
         way, check whether it is satisfied.}\label{f:flowchart}
\end{figure}

All scenarios discussed above and their corresponding sum rules have been 
summarised in figure~\ref{f:flowchart}. Related scenarios have been connected: 
if one starts at a given scenario, one should follow the arrows downwards to 
arrive at the underlying hypotheses. When we have determined the values of the 
RGIs from experimental data, we can test whether the listed scenarios are 
consistent with the data. One should proceed as follows: to test a hypothesis, 
check the validity of the sum rules in the corresponding box. Then check the 
validity of the sum rules in all boxes one encounters by following the arrows 
all the way down. If all these sum rules are satisfied, the hypothesis is 
consistent with the experimental data (as far as our sum rules are concerned).

\clearpage

\subsection{Model-specific sum rules}
Until now we have only considered hypotheses concerned with relations between 
the running parameters of the (p)MSSM. These hypotheses do not refer to any 
model-specific parameters. However, we can find additional sum rules for 
certain models because the soft masses are related by only a few parameters.
For example, in MGM the gaugino and sfermion masses are determined by the gauge
couplings and a single parameter $B$; see equations 
\eqref{eq:MGMgauginos}--\eqref{eq:MGMsfermions}. Furthermore, the question 
whether the messenger scale equals the gauge-coupling unification scale only 
makes sense if we consider gauge-mediation models. Therefore we consider 
model-specific sum rules separately in this section.

In the following, we will look for sum rules for GGM and AMSB that do not 
follow from the general hypotheses we have discussed above. Such sum rules will
generically be referred to as ``model-specific sum rules''. The sum rules for
GGM can also be found in \cite{Carena2}. To our knowledge no sum rules for AMSB
models have been presented in the literature. We will not discuss mSUGRA, 
because for our purposes the mSUGRA spectrum is completely characterised by 
simultaneous scalar-mass, gaugino-mass and gauge-coupling unification. 
So, there are no mSUGRA-specific sum rules. 

\subsubsection{General gauge mediation}
In section~\ref{s:susy-GGM} the RG boundary conditions for GGM are given in 
terms of 11 model parameters at the messenger scale $M$: $\delta_{u,d}$ and 
$B_a,A_a,g_a(M)$ for $a=1,2,3$ (we take $\zeta=0$). If we insert these RG boundary conditions into 
the RGIs, we immediately find \mbox{$D_{B_{13}}=D_{L_{13}}=0$,} as expected 
from flavour-universality. In addition we get {\it one model-specific sum 
rule}:
\begin{equation}
D_{\chi_1}=0. \label{eq:GGMmodelspecificsumrule}
\end{equation}

\noindent The remaining RGIs have the following values:
\begin{IEEEeqnarray}{rCl}\label{eq:GGMRGIs}
D_{Y_{13H}} &=& -\frac{10}{13}\left(\delta_u-\delta_d\right), 
                \IEEEyessubnumber\\
D_Z &=& -2\delta_d, \IEEEyessubnumber\\
I_{Y_\alpha} &=& g_1^{-2}\left(\delta_u-\delta_d\right), \IEEEyessubnumber\\
I_{B_a} &=& B_a \qquad (a=1,2,3), \IEEEyessubnumber\\
I_{M_1} &=& g_1^4\left(B_1^2 + \frac{33}{10}A_1\right), \IEEEyessubnumber\\
I_{M_2} &=& g_2^4\left(B_2^2 + \frac12A_2\right), \IEEEyessubnumber\\
I_{M_3} &=& g_3^4\left(B_3^2 - \frac32A_3\right), \IEEEyessubnumber\\
I_{g_2} &=& g_1^{-2} -\frac{33}{5}g_2^{-2}, \IEEEyessubnumber\\
I_{g_3} &=& g_1^{-2} + \frac{11}{5}g_3^{-2}, \IEEEyessubnumber
\end{IEEEeqnarray}
where the gauge couplings are understood to be evaluated at the messenger 
scale. This amounts to eleven equations with eleven unknowns, hence no 
additional sum rules can be constructed. Note that we can again verify 
non-universality in the Higgs sector using equations 
\eqref{eq:sumrule5a.4}--\eqref{eq:sumrule5a.5}.

\paragraph{Gauge-coupling unification at the messenger scale.}
If both gauge-coupling unification and GGM are compatible with experimental 
data, we may ask ourselves if the messenger scale equals the scale of 
gauge-coupling unification. If we insert $g_1=g_2=g_3=g$ into equations 
\eqref{eq:GGMRGIs}, we get eleven equations with nine unknowns.
Hence we can make two more sum rules. This includes equation 
\eqref{eq:sumrule1} for gauge-coupling unification. Hence, there is only 
{\it one additional model-specific sum rule}:
\begin{equation}
I_{Y_\alpha} = \frac{13}{56}I_{g_2}D_{Y_{13H}} \label{eq:messengergcsumrule}.
\end{equation}
Note that this sum rule happens to be identical to equation 
\eqref{eq:sumrule1+5}, which was constructed for testing a completely different
concept: simultaneous scalar-mass and gauge-coupling unification 
(i.e.~scenario~15). Therefore we have marked this sum rule as being 
model-specific. Also note that this sum rule becomes 
redundant in the case of universal Higgs masses, 
since $I_{Y_\alpha}=D_{Y_{13H}}=0$ in that case. 

\paragraph{Minimal gauge mediation.}
Recall that MGM is a GGM model restricted to a subset of the GGM parameter 
space defined by $A_a=A$, $B_a=B$ and $A=2B^2$. Inserting this into the RGI 
values \eqref{eq:GGMRGIs} of GGM, we find eleven non-vanishing RGIs that depend
on six parameters. Hence we can construct five sum rules. These include 
\eqref{eq:sumrule3.1}--\eqref{eq:sumrule3.2} for $M_a/g_a^2$ unification. 
There are {\it three additional model-specific sum rules}:\footnote{Note that 
in MGM, we can safely divide by $I_{M_a}$: if one of the $I_{M_a}$ vanished, 
then $B=0$ and the gaugino masses would vanish at the messenger scale. 
Their $\beta$-functions, being proportional to the gaugino masses, would vanish
as well. Then at one-loop order gauginos would be massless at all scales (only 
through two-loop effects the masses will be non-vanishing). In that case we 
would have observed them already. Thus the $I_{M_a}$ cannot vanish.}
\begin{IEEEeqnarray}{rCl}
0 &=& I_{Y_\alpha} + \frac{13}{10}D_{Y_{13H}}I_{B_1}
      \sqrt{\frac{38}{5I_{M_1}}}~,\label{eq:MGMsumrule1}\\
0 &=& I_{B_1}\sqrt{\frac{38}{5I_{M_1}}} 
      - \frac{33}{5}I_{B_1}\sqrt{\frac{2}{I_{M_2}}} - I_{g_2}, \\
0 &=& I_{B_1}\sqrt{\frac{38}{5I_{M_1}}} 
      + \frac{11}{5}I_{B_1}\sqrt{\frac{-2}{I_{M_3}}} - I_{g_3}.
\end{IEEEeqnarray}
Note that \eqref{eq:MGMsumrule1} becomes redundant in the case of universal 
Higgs masses.

\subsubsection{Anomaly mediation.}
The RG boundary conditions for AMSB are given in 
section~\ref{s:susy-AMSB} in terms of 4 model parameters at the scale
$M_{_\text{AMSB}}$ where supersymmetry breaking occurs: $m_{3/2}$ and $g_a(M_{_\text{AMSB}})$ for $a=1,2,3$. If we insert 
these RG boundary conditions into the RGIs, we immediately find 
{\it nine model-specific sum rules}:
\begin{equation}
D_{B_{13}} = D_{L_{13}} = D_{\chi_1} = D_{Y_{13H}} = D_Z = I_{Y_\alpha} 
= I_{M_1} = I_{M_2} = I_{M_3} = 0.
\end{equation}
Note that $D_{B_{13}}$ and $D_{L_{13}}$ vanish although the sfermion masses are
not flavour-universal! The non-vanishing RGIs have the values:
\begin{IEEEeqnarray}{rCl}\label{eq:AMSBRGI}
I_{B_1} &=& \frac{33}{5}\frac{m_{3/2}}{16\pi^2}, 
            \IEEEyessubnumber\label{eq:AMSBRGI1}\\
I_{B_2} &=& \frac{m_{3/2}}{16\pi^2}, \IEEEyessubnumber\label{eq:AMSBRGI2}\\
I_{B_3} &=& -3\frac{m_{3/2}}{16\pi^2}, \IEEEyessubnumber\label{eq:AMSBRGI3}\\
I_{g_2} &=& g_1^{-2} - \frac{33}{5}g_2^{-2}, 
            \IEEEyessubnumber\label{eq:AMSBRGI4}\\
I_{g_3} &=& g_1^{-2} + \frac{11}{5}g_3^{-2}, 
            \IEEEyessubnumber\label{eq:AMSBRGI5}
\end{IEEEeqnarray}
where the gauge couplings should be evaluated at the scale of supersymmetry
breaking. This amounts to five equations with four unknowns, but we can do 
better: equations \eqref{eq:AMSBRGI1}--\eqref{eq:AMSBRGI3} constitute three 
equations with one unknown. This yields {\it another two model-specific sum 
rules}:
\begin{align}
0 &= I_{B_1} - \frac{33}{5}I_{B_2}, \\
0 &= I_{B_1} + \frac{11}{5}I_{B_3}.
\end{align}

\paragraph{Gauge-coupling unification at the scale of supersymmetry breaking.}
If both AMSB and gauge-coupling unification turn out to be consistent with 
experimental data, we may ask ourselves whether supersymmetry breaking occurs 
at the scale of gauge-coupling unification. In that case we should insert 
$g_1=g_2=g_2=g$ into \eqref{eq:AMSBRGI}. But this will only affect equations 
\eqref{eq:AMSBRGI4} and \eqref{eq:AMSBRGI5}, which we have not used to make the
above sum rules. This amounts to two equations with only one parameter, so we 
get one more sum rule. This must be the sum rule \eqref{eq:sumrule1} for 
gauge-coupling unification, hence there are no sum rules that specifically test
whether gauge-coupling unification occurs at the scale of supersymmetry 
breaking.

\paragraph{Minimal anomaly mediation.}
Recall that in minimal AMSB, a universal additional term $m_0^2$ is added to 
the soft scalar masses.
If we insert this into the RGI expressions, we immediately find {\it five 
model-specific sum rules}:
\begin{equation}
D_{B_{13}} = D_{L_{13}} = D_{Y_{13H}} = D_Z = I_{Y_\alpha} = 0. \\
\end{equation}
The non-vanishing RGIs have the values:
\begin{IEEEeqnarray}{rCl}
D_{\chi_1} &=& 5m_0^2, \IEEEyessubnumber\\
I_{B_1} &=& \frac{33}{5}\frac{m_{3/2}}{16\pi^2}, \IEEEyessubnumber\\
I_{B_2} &=& \frac{m_{3/2}}{16\pi^2}, \IEEEyessubnumber\\
I_{B_3} &=& -3\frac{m_{3/2}}{16\pi^2}, \IEEEyessubnumber\\
I_{M_1} &=& \frac{33}{8}m_0^2, \IEEEyessubnumber\\
I_{M_2} &=& \frac58m_0^2, \IEEEyessubnumber\\
I_{M_3} &=& -\frac{15}{16}m_0^2, \IEEEyessubnumber\\
I_{g_2} &=& g_1^{-2} - \frac{33}{5}g_2^{-2}, 
            \IEEEyessubnumber\label{eq:mAMSBRGI1}\\
I_{g_3} &=& g_1^{-2} + \frac{11}{5}g_3^{-2}, 
            \IEEEyessubnumber\label{eq:mAMSBRGI2}
\end{IEEEeqnarray}
where again the gauge couplings should be evaluated at the scale of 
supersymmetry breaking. This adds up to nine equations with five unknowns, so 
we expect to find four additional sum rules. However, if we leave out equations
\eqref{eq:mAMSBRGI1}-\eqref{eq:mAMSBRGI2}, we are left with seven equations 
with only two unknowns. This yields {\it another five model-specific sum 
rules}:
\begin{align}
0 &= I_{B_1} - \frac{33}{5}I_{B_2}, \label{eq:mAMSB1}\\
0 &= I_{B_1} + \frac{11}{5}I_{B_3}, \label{eq:mAMSB2}\\
0 &= D_{\chi_1} - \frac{40}{33}I_{M_1}, \label{eq:mAMSB3}\\
0 &= D_{\chi_1} - 8I_{M_2}, \label{eq:mAMSB4}\\
0 &= D_{\chi_1} + \frac{16}{3}I_{M_3}. \label{eq:mAMSB5}
\end{align}
Here equations \eqref{eq:mAMSB1}-\eqref{eq:mAMSB2} also hold for AMSB.
Equations \eqref{eq:mAMSB3}-\eqref{eq:mAMSB5} are automatically satisfied in 
AMSB because $D_{\chi_1}$ and $I_{M_a}$ vanish.

\subsection{Discussion}\label{s:discussion}
In this section, we have found (a) sum rules that test general properties of 
the RG boundary conditions and (b) sum rules that test the consistency of 
specific model spectra. Comparing both sets of sum rules will help us to 
determine how good the sum rules are at distinguishing between several 
properties and model spectra. 

\paragraph{Sum-rule ambiguities and how to eliminate them.}
In the sum rules we observe the following ambiguities:
\begin{itemize}
\item If the sum rules \eqref{eq:sumrule1} for gauge-coupling unification and 
      \eqref{eq:sumrule3.1}--\eqref{eq:sumrule3.2} for $M_a/g_a^2$ unification 
      are both satisfied, then the sum rule \eqref{eq:sumrule2} for 
      gaugino-mass unification is automatically satisfied. But gaugino-mass 
      unification is implied by gauge-coupling unification and $M_a/g_a^2$ 
      unification only if both unifications occur at the same scale! Hence, if 
      \eqref{eq:sumrule1}, \eqref{eq:sumrule3.1} and \eqref{eq:sumrule3.2} are 
      satisfied by experimental data, then we cannot determine unambiguously 
      whether the gaugino masses unify. At this point, we should use the 
      bottom-up method to examine the running of the parameters. Then we could 
      see whether the unification scales are the same.
\item Equation \eqref{eq:sumrule1+5} checks whether scalar masses and gauge 
      couplings unify at the same scale. Equation \eqref{eq:messengergcsumrule}
      checks whether the gauge couplings unify at the messenger scale in GGM. 
      Yet these sum rules happen to be the same. However, this does not mean we
      cannot distinguish between these two scenarios. The former scenario also 
      requires that the sum rules 
      \eqref{eq:sumrule5a.1}--\eqref{eq:sumrule5a.3} for scalar-mass 
      unification are valid. In the latter scenario, these sum rules are not 
      satisfied. Thus the double role of \eqref{eq:sumrule1+5} poses no 
      problem.
\item In AMSB and mAMSB, the sum rules 
      \eqref{eq:sumrule4.1}--\eqref{eq:sumrule4.2} for flavour-universality are
      satisfied, although the sfermion masses in these models are clearly
      non-universal. Fortunately, (m)AMSB has a lot more sum rules, which could
      help discern these models from flavour-universal ones. For example, the 
      vanishing of $D_{Y_{13H}}$, $D_Z$ and $I_{Y_\alpha}$ is typical for 
      (m)AMSB. Equations \eqref{eq:sumrule5a.1}--\eqref{eq:sumrule5a.2} then 
      help us discern (m)AMSB from scalar-mass unification with universal Higgs
      masses. Again, satisfying a single sum rule may be ambiguous, but other 
      sum rules eliminate this ambiguity.
\item Because $D_{Y_{13H}}$ and $I_{Y_\alpha}$ vanish in (m)AMSB, the sum rule
      \eqref{eq:sumrule1+5} for simultaneous gauge-coupling and scalar-mass 
      unification is automatically satisfied. However, the sum rules for 
      gauge-coupling unification and scalar-mass unification again help us 
      distinguish between both scenarios.
\item The vanishing of $I_{M_a}$ in AMSB and the sum rules 
      \eqref{eq:mAMSB3}--\eqref{eq:mAMSB5} of mAMSB both imply that the sum 
      rule \eqref{eq:sumrule2+5} for simultaneous scalar-mass and gaugino-mass 
      unification is satisfied. However, the sum rules for scalar-mass 
      unification and gaugino-mass unification help us distinguish between both
      scenarios.
\end{itemize}

If we only consider the spectrum properties and breaking mechanisms that we 
discussed in this section, our sum rules work surprisingly well. Many of the 
sum rules are not unambiguous by themselves, but in most cases the other sum 
rules remove the ambiguity. Only when the data are consistent with both 
gauge-coupling unification and $M_a/g_a^2$ unification, we have to resort to 
other methods (such as the bottom-up method) to determine whether the gaugino 
masses also unify (or equivalently, whether both unifications occur at the same
scale).

Of course, it is possible that a new supersymmetry-breaking model is concocted 
in the future, and that some of its corresponding sum rules introduce similar 
ambiguities. These may or may not be resolved by other sum rules.
Therefore, we should keep in mind that if the sum rules of a model or 
hypothesis are satisfied, this is not a confirmation that this model or 
hypothesis is correct. \emph{The true power of our sum rules is their 
falsifying power: the failure to satisfy just one sum rule implies that the 
corresponding hypothesis or model is incorrect.} 
This falsifying power depends on how precisely the parameters entering the RGIs
have been measured: the smaller the errors are, the more powerful the rejection becomes.
\vskip3ex
\noindent Now that we have an idea of the quality of the RGI sum rules, we can 
finally examine the advantages and limitations of the RGI method.

\paragraph{Advantages of RGIs}
\begin{itemize}
\item The RGI method requires less input than the other methods we have 
      discussed. We only need the values of all soft masses and gauge couplings
      at one scale. These are sufficient to reconstruct the values of the RGIs 
      in equations~\eqref{eq:db13}--\eqref{eq:ig3}. In contrast to the bottom-up method, we do not
      need the values of the Yukawa couplings and soft trilinear couplings 
      because we could not use them anyway. Also, the value of the 
      new-physics threshold does not have to be known.
\item The RGI method is very simple: it is entirely algebraical and does not 
      require the numerical integration of renormalisation group equations. 
      Therefore it avoids the complicated propagation of errors between the 
      collider scale and the new-physics threshold. Also, it is not as 
      time-consuming as the top-down method. 
\item If no unification of any sort is observed, the bottom-up method provides us
      with no clues about new physics, whereas the RGI method may still be helpful. Suppose for example
      that non-minimal General Gauge Mediation is a good model of Nature, so that no unification occurs.
      Then the bottom-up method does not provide any hints on where to look for new-physics
      thresholds. We could check whether the sfermion masses match the pattern~\eqref{eq:GGMsfermionmasses}
      at any scale, but that would take a lot of time. However, if we apply the RGI method we only need to
      check the three sum rules~\eqref{eq:sumrule4.1}, \eqref{eq:sumrule4.2} and \eqref{eq:GGMmodelspecificsumrule}
      to see whether this bottom-up scan is worth the effort.
\item As long as just a few of the relevant soft masses have been measured 
      experimentally, the sum rules can be exploited as a fast means of 
      identifying theoretically interesting regions in the remaining parameter 
      space (e.g.~regions with specific unifications). Here we assume that the particles corresponding to
      the masses have been identified correctly. The precise procedure for reconstructing the MSSM parameters from experimental data is beyond
      the scope of our study and will be investigated elsewhere. 
\end{itemize}

\paragraph{Limitations and challenges of the RGI method}
\begin{itemize}
\item As we mentioned before, RG invariance only holds up to a certain loop 
      level. The RGIs in~\eqref{eq:mssmrgis} have been determined using the 
      one-loop RG equations. Higher-order loop effects will certainly spoil RG
      invariance. We could of course try to find RGIs for the MSSM at a higher 
      loop order. But already at the two-loop level the RG equations for the 
      MSSM (see e.g.\ \cite{twoloopRGEs}) are too complicated to retain the 
      simplicity of this method, if it is possible to find RGIs at all.

      However, the relevant question is to what extent we should worry about 
      this approximate RG invariance. It has been demonstrated in 
      \cite{Carena1} for the MSSM that even if we have determined all soft masses at the collider scale 
      with a universal experimental uncertainty as low as 1\%, two-loop contributions to the RGIs are smaller than 
      the experimental errors in the RGIs. Thus for all practical purposes we can safely treat the one-loop RGIs as true 
      invariants. It should be noted that if we were to apply the RGI method to an EFT other than 
      the MSSM, we should again study the size of 2-loop contributions to the RGIs before we 
      can neglect them.
\item It may seem like the RGI method magically reduces the uncertainties of 
      the running parameters, compared to RG-evolved parameters. However, we 
      have paid a price for this reduction, namely \emph{information}. We can 
      directly see this from equations~\eqref{eq:db13}--\eqref{eq:ig3}: we started with 18 running
      parameters (12 scalar masses, 3 gaugino masses and 3 gauge couplings) and
      have reduced them to only 14 invariants.

      We can easily understand why we have to give up information to gain 
      smaller errors. Consider for example the RG equations for $\msqa$ and 
      $\msqc$ (see appendix~\ref{a:mssmrges} for the definitions of 
      $D_Y,X_t,X_b$):
      \begin{IEEEeqnarray}{rCl}
        16\pi^2\frac{\d \msqa}{\d t} &=& -\frac{2}{15}g_1^2M_1^2 - 6g_2^2M_2^2 
        - \frac{32}{3}g_3^2M_3^2 + \frac15g_1^2D_Y, \IEEEyessubnumber\\
        16\pi^2\frac{\d \msqc}{\d t} &=& X_t + X_b -\frac{2}{15}g_1^2M_1^2 
        - 6g_2^2M_2^2 - \frac{32}{3}g_3^2M_3^2 + \frac15g_1^2D_Y. \qquad\quad 
      \IEEEyessubnumber
      \end{IEEEeqnarray}
      Note that in the RG equations of all soft masses, dependence on the 
      gaugino mass $M_2$ occurs only as terms proportional to $g_2^2M_2^2$.
      Hence we can eliminate the $M_2$ dependence by taking suitable linear
      combinations of MSSM parameters. For example, the RG equation for the 
      quantity $\msqa-\msqc$ (which occurs in $D_{B_{13}}$) does not depend on 
      $M_2$ any more, so its experimental uncertainty will spread less under RG
      flow. However, in this process we have thrown away information about the 
      value of $\msqa+\msqc$. Thus we have to reduce the number of independent
      quantities to reduce the spread of uncertainties under RG flow.

      This may become a limitation of the RGI method in the following sense. 
      A minimal model such as mSUGRA, with only three parameters that govern 
      the soft masses plus gauge couplings at the GUT scale 
      ($m_0,M_{1/2},g\equiv g_a(M_\text{GUT})$), allows us to construct sum 
      rules because we have more RGIs than mSUGRA has parameters. However, if 
      we have a not-so-minimal model with (say) 15 parameters that determine 
      the high-scale spectrum, we do not have enough RGIs to make any sum 
      rules.\footnote{That is, unless a subset of $n$ RGIs accidentally depends
      on less than $n$ model parameters.} Hence, despite the simplicity of the 
      method, we are still limited to models with few parameters.
\item The applicability of the RGI method to the study of supersymmetry 
      breaking depends crucially on the assumption that the MSSM 
      renormalisation group equations are valid all the way up to the scale of 
      supersymmetry breaking. But suppose that in Nature a new field $\Phi$ 
      (or possibly more than one) enters the theory at a high scale $\mu_\Phi$ 
      that is not the scale of supersymmetry breaking; instead supersymmetry is
      broken at an even higher scale $\mu_\text{SUSY}$. Then at $\mu_\Phi$ the 
      physical RG trajectories of the running parameters will be deflected from
      their MSSM trajectories. Thus we might mistakingly see gaugino-mass 
      unification where it is absent, or vice versa. Hence, if we want to study
      supersymmetry breaking directly from RGIs, we have to assume that new 
      physics, if present, does not alter the one-loop RG equations for the 
      MSSM up to the scale of supersymmetry breaking.
\item In order to make conclusive statements based on sum rules, the values
      of all RGIs should be reconstructed. To achieve that, all soft masses and
      gauge couplings need to be known at one energy scale. This may prove 
      difficult in practice. First of all, due to mixing effects the gauge 
      eigenstates do not always correspond to the mass eigenstates. 
      Reconstructing the soft masses from measured pole masses will introduce 
      additional uncertainties. Furthermore, determining all soft masses and 
      gauge couplings is one thing, but determining all of them \emph{at the 
      same energy scale} may prove challenging. Note however that the bottom-up
      method also suffers from these complications. 
\item If we find some sort of unification to be consistent with experimental
      data, the RGI method does not provide us with any information on the 
      value of the new-physics threshold. In order to determine the value of the 
      scale of new physics, we need to resort to other methods such as the
      bottom-up method.
\end{itemize}

\section{Conclusions and Outlook}\label{s:conclusions}

We advocate to employ sum rules in terms of renormalisation group 
invariants as a simple yet powerful method to probe high-scale physics in 
lower-energy experiments. This method has been worked out in detail 
for the study of supersymmetry-breaking mechanisms in the context of the 
Minimal Supersymmetric Standard Model. It has been argued that important clues 
about the supersymmetry-breaking mechanism are to be found in patterns between 
the high-scale soft-supersymmetry-breaking parameters. The renormalisation 
group is the prime tool to extract such information on the high-scale spectrum 
from lower-energy data. Several methods have been discussed to do this and a 
new strategy has been proposed to make effective use of renormalisation group 
invariants. Assuming that the Minimal Supersymmetric Standard Model is an 
appropriate effective field theory beyond the Standard Model, 
a model-independent set of renormalisation-group-invariant sum rules has been 
constructed that test properties that are common in supersymmetry-breaking 
models, such as unifications and flavour-universality. If a certain property is
realised in Nature, all corresponding sum rules must be satisfied. Since none 
of these sum rules refer to any parameters that are specific to some 
supersymmetry-breaking mechanism, they are useful regardless of the way 
supersymmetry has been broken in Nature. 

In addition, sum rules that are tailor-made for testing specific 
supersymmetry-breaking mechanisms have been considered. Their primary use was 
to determine the effectiveness and ambiguities associated with the 
model-independent sum rules. It was found that some sum rules do not provide 
unambiguous checks by themselves; however, in almost all cases other sum rules 
lift the ambiguity. Hence, for the currently known supersymmetry-breaking 
mechanisms the proposed model-independent sum rules are
surprisingly effective. In the exceptional case when they are not, one may have
to resort to other methods to resolve the ambiguity, such as a bottom-up 
analysis. It is possible that new breaking mechanisms will be proposed in the 
future, and that their corresponding sum rules introduce new ambiguities. 
Therefore, it should be kept in mind that the main strength of invariant sum 
rules is their falsifying power. If we are able to determine all soft masses 
and gauge couplings, the compatibility of the sum rules with experimental data 
will put severe constraints on any realistic model of supersymmetry breaking . 

It is possible that the next effective field theory beyond the Standard Model 
is not the Minimal Supersymmetric Standard Model. It may as well be a 
non-minimal supersymmetric extension of the Standard Model, or even a 
non-supersymmetric theory. Nevertheless, the proposed scheme for probing 
high-scale properties of running parameters may be applied just the same. 
In order to perform an analogous study, one needs to determine the particle 
content, interactions and $\beta$-functions of the appropriate effective field 
theory. Then one should determine all independent renormalisation group 
invariants for this effective field theory and construct sum rules in a way 
similar to what has been worked out in this study. However, there is a large 
amount of structure in the $\beta$-functions of the Minimal Supersymmetric 
Standard Model in view of the limited number of combinations in which running 
parameters appear in them. So, an interesting topic is to determine the form of
the (one-loop) $\beta$-functions for a theory that is more general than the 
Minimal Supersymmetric Standard Model, and to see what renormalisation group 
invariants can be found for such a general theory. We leave these 
issues to future work.

Apart from establishing the validity of the Minimal Supersymmetric Standard 
Model, the main obstacle to using our invariant sum rules is the necessity of 
knowing all soft masses and gauge couplings at one scale. Therefore, an 
important topic for future study will be to determine how well this can be 
done and how the sum rules can be exploited as a fast means of identifying 
theoretically interesting regions in as yet unconstrained parameter space. 
Another important issue is to find out how the soft mass parameters can be 
reconstructed from the mass eigenstates of the sparticles. 

\section*{Acknowledgments}

We thank Jari Laamanen and Irene Niessen for useful discussions. 
This work has been supported in part by the Foundation for Fundamental
Research of Matter (FOM), program~104 ``Theoretical Particle Physics in
the Era of the LHC". 

\newpage
\appendix

\section{One-loop RG equations for the MSSM}\label{a:mssmrges}
In this appendix we give the renormalisation group equations of the MSSM that 
have been used in this study. They have been taken from \cite{martin} and are 
one-loop equations that have been simplified by the assumptions for the pMSSM (see section~\ref{s:susy-constraints}).
For general two-loop RG equations, see e.g.\ \cite{twoloopRGEs}. It is 
convenient to use the $\beta$-functions, which differ from their corresponding 
RG equations by a constant:
\begin{equation}
\beta(p) \equiv 16\pi^2\frac{\d p}{\d t}.
\end{equation}
Here $p$ is a running parameter and $t\equiv\log{\left(\mu/\mu_0\right)}$, 
where $\mu$ is the renormalisation scale and $\mu_0$ an (arbitrary) energy 
scale that makes the argument of the logarithm dimensionless. Under the 
approximations of the pMSSM we are left with the following running parameters:
\begin{IEEEeqnarray*}{ll}
g_a \qquad(a=1,2,3) & \text{Gauge couplings} \\
M_a \quad\ \:(a=1,2,3) & \text{Soft-supersymmetry-breaking gaugino masses} \\
\msq\,,\msu\,,\msd\,,\msl\,,\mse \quad & \text{Soft-supersymmetry-breaking 
                                               sfermion masses} \\
\mhu, \mhd & \text{Soft-supersymmetry-breaking Higgs mass parameters} \\
y_t,y_b,y_\tau & \text{Yukawa couplings for the third-generation (s)fermions}\\
A_t,A_b,A_\tau & \text{Soft-supersymmetry-breaking trilinear couplings for the}
\\  & \text{third-generation sfermions} \\
\mu & \text{Supersymmetry-respecting Higgs mixing parameter} \\
B & \text{Soft-supersymmetry-breaking Higgs mixing parameter}
\end{IEEEeqnarray*}
Here we use the soft Higgs mixing parameter $B=b/\mu$ rather than $b$ because 
its $\beta$-function is simpler. For the sfermion masses, we denote the first 
and third generation with a subscript 1 and 3 respectively.

The following notation is used for parameters that enter the RG equations 
through common combinations of Dynkin indices and quadratic Casimir invariants:
\begin{equation}
b_a = \left(\frac{33}{5}, 1, -3\right) \qquad \mbox{for\ } a=1,2,3.
\end{equation}
It is also convenient to define the following combination of running 
parameters, which appears in the RG equations of the sfermion masses:
\begin{equation}
D_Y \equiv \text{Tr}\left(Ym^2\right) 
    = \sum_\text{gen} \left(\msq-2\msu+\msd-\msl+\mse\right) + \mhu-\mhd.
\end{equation}
Here the trace runs over all chiral multiplets and the sum runs over the three 
sfermion generations. Note that $D_Y$ is often called $S$ in the literature.
Furthermore, we define the useful combinations:
\begin{IEEEeqnarray}{lCl}
X_t &=& 2|y_t|^2\left(\mhu + \msqc + \msuc + |A_t|^2\right), 
        \IEEEyessubnumber\\
X_b &=& 2|y_b|^2\left(\mhd + \msqc + \msdc + |A_b|^2\right), 
        \IEEEyessubnumber\\
X_\tau &=& 2|y_\tau|^2\left(\mhd + \mslc + \msec + |A_\tau|^2\right). 
           \IEEEyessubnumber
\end{IEEEeqnarray}
Then the resulting $\beta$-functions for the MSSM are:
\begin{IEEEeqnarray}{lCl}
\beta(g_a) &=& b_ag_a^3 \qquad (a=1,2,3), \IEEEyessubnumber\\
\beta(M_a) &=& 2b_ag_a^2M_a \qquad (a=1,2,3), \IEEEyessubnumber\\
\beta(m_{\tilde{Q}_{1,2}}^2) &=& -\frac{2}{15}g_1^2M_1^2 - 6g_2^2M_2^2 - \frac{32}{3}g_3^2M_3^2 + \frac{1}{5}g_1^2D_Y, \IEEEyessubnumber\\
\beta(m_{\tilde{\bar{u}}_{1,2}}^2) &=& -\frac{32}{15}g_1^2M_1^2 - \frac{32}{3}g_3^2M_3^2 - \frac{4}{5}g_1^2D_Y, \IEEEyessubnumber\\
\beta(m_{\tilde{\bar{d}}_{1,2}}^2) &=& -\frac{8}{15}g_1^2M_1^2 - \frac{32}{3}g_3^2M_3^2 + \frac{2}{5}g_1^2D_Y, \IEEEyessubnumber\\
\beta(m_{\tilde{L}_{1,2}}^2) &=& -\frac{6}{5}g_1^2M_1^2 - 6g_2^2M_2^2 - \frac{3}{5}g_1^2D_Y, \IEEEyessubnumber\\
\beta(m_{\tilde{\bar{e}}_{1,2}}^2) &=& -\frac{24}{5}g_1^2M_1^2 + \frac{6}{5}g_1^2D_Y, \IEEEyessubnumber\\
\beta(m_{\tilde{Q}_3}^2) &=& X_t + X_b - \frac{2}{15}g_1^2M_1^2 - 6g_2^2M_2^2 - \frac{32}{3}g_3^2M_3^2 + \frac{1}{5}g_1^2D_Y, \IEEEyessubnumber\\
\beta(m_{\tilde{\bar{u}}_3}^2) &=& 2X_t - \frac{32}{15}g_1^2M_1^2 - \frac{32}{3}g_3^2M_3^2 - \frac{4}{5}g_1^2D_Y, \IEEEyessubnumber\\
\beta(m_{\tilde{\bar{d}}_3}^2) &=& 2X_b - \frac{8}{15}g_1^2M_1^2 - \frac{32}{3}g_3^2M_3^2 + \frac{2}{5}g_1^2D_Y, \IEEEyessubnumber\\
\beta(m_{\tilde{L}_3}^2) &=& X_\tau - \frac{6}{5}g_1^2M_1^2 - 6g_2^2M_2^2 - \frac{3}{5}g_1^2D_Y, \IEEEyessubnumber\\
\beta(m_{\tilde{\bar{e}}_3}^2) &=& 2X_\tau - \frac{24}{5}g_1^2M_1^2 + \frac{6}{5}g_1^2D_Y, \IEEEyessubnumber\\
\beta(m_{H_u}^2) &=& 3X_t - \frac{6}{5}g_1^2M_1^2 - 6g_2^2M_2^2 + \frac{3}{5}g_1^2D_Y, \IEEEyessubnumber\\
\beta(m_{H_d}^2) &=& 3X_b + X_\tau - \frac{6}{5}g_1^2M_1^2 - 6g_2^2M_2^2 - \frac{3}{5}g_1^2D_Y, \IEEEyessubnumber\\
\beta(y_t) &=& y_t\left[6|y_t|^2 + |y_b|^2 - \frac{13}{15}g_1^2 - 3g_2^2 - \frac{16}{3}g_3^2\right], \IEEEyessubnumber\\
\beta(y_b) &=& y_b\left[6|y_b|^2 + |y_t|^2 + |y_\tau|^2 - \frac{7}{15}g_1^2 - 3g_2^2 - \frac{16}{3}g_3^2\right], \IEEEyessubnumber\\
\beta(y_\tau) &=& y_\tau\left[4|y_\tau|^2 + 3|y_b|^2 - \frac{9}{5}g_1^2 - 3g_2^2\right], \IEEEyessubnumber\\
\beta(\mu) &=& \mu\left[3|y_t|^2 + 3|y_b|^2 + |y_\tau|^2 - \frac{3}{5}g_1^2 - 3g_2^2\right], \IEEEyessubnumber\\
\beta(A_t) &=& 12A_t|y_t|^2 + 2A_b|y_b|^2 + \frac{26}{15}g_1^2M_1 + 6g_2^2M_2 + \frac{32}{3}g_3^2M_3, \IEEEyessubnumber\\
\beta(A_b) &=& 12A_b|y_b|^2 + 2A_t|y_t|^2 + 2A_\tau|y_\tau|^2 \nonumber\\
&&\qquad\qquad\qquad\qquad\quad + \frac{14}{15}g_1^2M_1 + 6g_2^2M_2 + \frac{32}{3}g_3^2M_3, \IEEEyessubnumber\\
\beta(A_\tau) &=& 8A_\tau|y_\tau|^2 + 6A_b|y_b|^2 + \frac{18}{5}g_1^2M_1 + 6g_2^2M_2, \IEEEyessubnumber\\
\beta(B) &=& 6A_t|y_t|^2 + 6A_b|y_b|^2 + 2A_\tau|y_\tau|^2 + \frac{6}{5}g_1^2M_1 + 6g_2^2M_2. \IEEEyessubnumber
\end{IEEEeqnarray}

\section{Deriving the one-loop RGIs for the MSSM}\label{a:mssmrgis}
In this appendix we will derive a maximal set of independent RGIs for the MSSM.
First we will determine invariants that contain the running parameters $\mu$ 
and $B=b/\mu$. We will see that there is only one independent RGI for each of 
them, making them useless for our study. Then we will argue that we are 
restricted to RGIs containing only soft masses and/or gauge couplings.
We will derive all of them systematically; our approach will be globally the 
same as in \cite{Carena1}, but using different arguments to show that we do 
indeed find all RGIs.

Let us consider the parameter $\mu$. The only $\beta$-function containing $\mu$
is that of $\mu$ itself. Note that we can write $\beta(\mu)$ more conveniently 
as:
\begin{equation}
\beta(\log{\mu}) = 3|y_t|^2 + 3|y_b|^2 + |y_\tau|^2 - \frac35g_1^2 - 3g_2^2.
\end{equation}
The only other $\beta$-functions containing terms linear in 
$|y_t|^2,|y_b|^2,|y_\tau|^2$ are those of the logarithms of the Yukawa 
couplings:
\begin{IEEEeqnarray}{rCl}
\beta(\log{y_t}) &=& 6|y_t|^2 + |y_b|^2 - \frac{13}{15}g_1^2 - 3g_2^2 
                     - \frac{16}{3}g_3^2, \IEEEyessubnumber\\
\beta(\log{y_b}) &=& |y_t|^2 + 6|y_b|^2 + |y_\tau|^2 - \frac{7}{15}g_1^2 
                     - 3g_2^2 - \frac{16}{3}g_3^2, \IEEEyessubnumber\\
\beta(\log{y_\tau}) &=& 3|y_b|^2 + 4|y_\tau|^2 - \frac95g_1^2 - 3g_2^2. 
                        \IEEEyessubnumber
\end{IEEEeqnarray}
The terms in the $\beta$-functions proportional to $g_a^2$ can be eliminated by
taking linear combinations with logarithms of gauge couplings, of which we can 
rewrite the $\beta$-functions as:
\begin{equation}
\beta(\log{g_a}) = b_ag_a^2 \qquad (a=1,2,3). \label{eq:loggaugebetafunctions}
\end{equation}
Hence, $\mu$ can only appear in an RGI through a linear combination of 
$\log{\mu}$, $\log{y_t}$, $\log{y_b}$, $\log{y_\tau}$, $\log{g_1}$, $\log{g_2}$
and $\log{g_3}$.\footnote{We could also include logarithms of gaugino masses in
these linear combinations, since their $\beta$-functions are also proportional 
to $g_a^2$. However, in a moment we will construct RGIs from the gauge 
couplings and gaugino masses only. Any RGI that contains both $\mu$ and the 
gaugino masses will be a function of those RGIs and the one we are constructing
now.} We have seven $\beta$-functions with six different terms to eliminate 
(namely terms linear in $|y_t|^2,|y_b|^2,|y_\tau|^2,g_1^2,g_2^2$ or $g_3^2$), 
so we can make one RG invariant linear combination of them. Using elementary 
linear algebra we find that the linear combination
\begin{align}
&-\frac{27}{61}\log{y_t} - \frac{21}{61}\log{y_b} - \frac{10}{61}\log{y_\tau} 
 + \log{\mu} - \frac{1}{61}\cdot\frac{73}{33}\log{g_1} \nonumber\\
&\qquad\qquad+ \frac{9}{61}\log{g_2} + \frac{1}{61}\cdot\frac{256}{3}\log{g_3} 
 \nonumber \\
=& \log{\left(\mu\left[\frac{g_2^9\,g_3^{256/3}}
   {y_t^{27}y_b^{21}y_\tau^{10}g_1^{73/33}}\right]^{1/61}\right)}
\end{align}
has a vanishing $\beta$-function. Thus we can choose the only independent RGI 
containing $\mu$ to be
\begin{equation}
I_2 \equiv \mu\left[\frac{g_2^9\,g_3^{256/3}}
           {y_t^{27}y_b^{21}y_\tau^{10}g_1^{73/33}}\right]^{1/61},
\end{equation}
using the notation of \cite{Demir}.
To summarise, we have found a set of independent RGIs containing $\mu$ (in this
case only one) by considering what terms in the MSSM $\beta$-functions could 
cancel each other. This will be our general strategy for finding all RGIs of 
the MSSM, because the running parameters only enter the $\beta$-functions in a 
very limited number of combinations (e.g.\ the soft scalar masses only appear 
in the linear combinations $D_Y$, $X_t$, $X_b$ and $X_\tau$).

Now we turn to the parameter $B$. It does not appear in any of the MSSM 
$\beta$-functions itself. Its $\beta$-function contains only terms linear in 
$A_t|y_t|^2$, $A_b|y_b|^2$, $A_\tau|y_\tau|^2$, $g_1^2M_1$, $g_2^2M_2$ and 
$g_3^2M_3$. The $\beta$-functions of $A_t$, $A_b$, $A_\tau$, $M_1$, $M_2$ and 
$M_3$ also contain only these terms, so $B$ should always appear in RGIs in a 
linear combination of these parameters. This gives us seven $\beta$-functions 
with six different terms to eliminate, so again we can make one RG-invariant 
linear combination. Using elementary linear algebra this combination is found 
to be
\begin{equation}
I_4 \equiv B - \frac{27}{61}A_t -\frac{21}{61}A_b -\frac{10}{61}A_\tau 
           - \frac{256}{183}M_3 -\frac{9}{61}M_2 +\frac{73}{2013}M_1.
\end{equation}

Indeed we have found only one independent RGI containing $\mu$ and one 
containing~$B$. As was argued in section~\ref{s:susy-rgiliterature}, RGIs are only useful as long as their 
constituent running parameters also appear in other RGIs. This is not the case 
for $I_2$ and $I_4$, so we are restricted to RGIs that contain neither $\mu$ 
nor $B$. But in the above procedure, we needed their $\beta$-functions to 
eliminate the $|y_i|^2$ and $A_i|y_i|^2$ dependence respectively from the 
$\beta$-function of the RGI under construction. If we wish to construct RGIs 
containing the Yukawa couplings without using $\mu$, we have to eliminate three
different $|y_i|^2$ terms using three $\beta$-functions, so we cannot make any 
RG-invariant combinations. Similarly, we cannot make any RGIs containing the 
soft trilinear couplings without using $B$, because we have to eliminate three 
different $A_i|y_i|^2$ terms using three $\beta$-functions.

Thus, if we want to construct RGIs without using $\mu$ and $B$, we cannot use 
the Yukawa and soft trilinear couplings either: we do not have enough equations
to eliminate all terms from the $\beta$-function of the RGI under construction.
Therefore, from now on we will only consider RGIs that are functions of soft 
masses (fifteen parameters) and/or gauge couplings (three parameters).

Let us begin with RGIs constructed from the gauge couplings only. First we 
rewrite their $\beta$-functions into a more convenient form:
\begin{IEEEeqnarray}{rCl}
\beta(g_1^{-2}) &=& -\frac{66}{5}, \IEEEyessubnumber \\
\beta(g_2^{-2}) &=& -2, \IEEEyessubnumber \\
\beta(g_3^{-2}) &=& 6. \IEEEyessubnumber
\end{IEEEeqnarray}
This gives us three equations to eliminate a single term (namely a constant), 
hence we can make two independent RGIs out of them. In accordance with 
\cite{Carena1}, we choose them to be
\begin{align}
I_{g_2} &\equiv g_1^{-2} - \frac{33}{5}g_2^{-2}, \\
I_{g_3} &\equiv g_1^{-2} + \frac{11}{5}g_3^{-2}.
\end{align}
Now we turn to the gaugino masses. First we rewrite their $\beta$-functions as 
follows:
\begin{IEEEeqnarray}{rCl}
\beta(\log{M_1}) &=& \frac{66}{5}g_1^2, \IEEEyessubnumber \\
\beta(\log{M_2}) &=& 2g_2^2, \IEEEyessubnumber \\
\beta(\log{M_3}) &=& -6g_3^2. \IEEEyessubnumber
\end{IEEEeqnarray}
Together with \eqref{eq:loggaugebetafunctions} this gives six equations with 
three different terms (namely those proportional to $g_a^2$) to eliminate.
Hence, we get three new RGIs by taking linear combinations of $\log{M_a}$ and 
$\log{g_a}$:
\begin{IEEEeqnarray}{rCl}
0 &=& \beta(\log{M_1} - 2\log{g_1}) = \beta(\log{\frac{M_1}{g_1^2}}), 
      \IEEEyessubnumber \\
0 &=& \beta(\log{M_2} - 2\log{g_2}) = \beta(\log{\frac{M_2}{g_2^2}}), 
      \IEEEyessubnumber \\
0 &=& \beta(\log{M_3} - 2\log{g_3}) = \beta(\log{\frac{M_3}{g_3^2}}).
      \IEEEyessubnumber
\end{IEEEeqnarray}
Thus we can choose the three independent RGIs to be:
\begin{align}
I_{B_1} &\equiv \frac{M_1}{g_1^2}, \\
I_{B_2} &\equiv \frac{M_2}{g_2^2}, \\
I_{B_3} &\equiv \frac{M_3}{g_3^2}.
\end{align}
Now let us consider RGIs constructed solely from the twelve soft scalar masses.
First we eliminate the Yukawa terms $X_t$, $X_b$, $X_\tau$ and the gaugino-mass
terms $g_1^2M_1^2$, $g_2^2M_2^2$, $g_3^2M_3^2$ from the $\beta$-function.
Since we have to eliminate six terms using twelve equations, we can make six 
independent linear combinations of the soft scalar masses that have a 
$\beta$-function proportional to $g_1^2D_Y$. Then we can make linear 
combinations of these quantities such that five of them have a vanishing 
$\beta$-function and the sixth quantity still runs with $g_1^2D_Y$.
In accordance with \cite{Carena1}, we choose the five RGIs to be:\footnote{The 
notation used for the RGIs may look odd here. In \cite{Carena1} they are 
related to symmetries of the MSSM Lagrangian. In this context, the $D$-term 
$D_i$ of a charge $Q_i$ is defined as $D_i\equiv \text{Tr}(Q_im^2)$, with the 
trace running over all chiral multiplets. Then one should interpret 
$D_{B_{13}}$ as \mbox{$D_{B_1}-D_{B_3}$,} where the subscripts~1 and 3 mean 
that the trace is restricted to sfermions of the first and third generation 
respectively. See \cite{Carena1} for an explanation of the nomenclature for the
remaining RGIs.}
\begin{align}
D_{B_{13}} &\equiv 2\left(\msqa-\msqc\right) - \msua + \msuc - \msda + \msdc,\\
D_{L_{13}} &\equiv 2\left(\msla-\mslc\right) - \msea + \msec, \\
D_{\chi_1} &\equiv 3\left(3\msda-2\left(\msqa-\msla\right)-\msua\right)-\msea,
                   \\
D_{Y_{13H}} &\equiv \msqa-2\msua+\msda-\msla+\msea \nonumber\\
&-\frac{10}{13}\left(\msqc-2\msuc+\msdc-\mslc+\msec+\mhu-\mhd\right), \\
D_Z &\equiv 3\left(\msdc-\msda\right) + 2\left(\mslc-\mhd\right).
\end{align}
The sixth quantity, which runs with $g_1^2D_Y$, can be chosen to be $D_Y$ 
itself, because
\begin{equation}
\beta(D_Y) = \frac{66}{5}g_1^2D_Y.
\end{equation}
Note that $\log{D_Y}$ runs with $g_1^2$, so using 
\eqref{eq:loggaugebetafunctions} we find
\begin{equation}
\beta(\log{D_Y} - 2\log{g_1}) = \beta(\log{\frac{D_Y}{g_1^2}}) = 0.
\end{equation}
This gives us another independent RGI:
\begin{equation}
I_{Y_\alpha} \equiv \frac{D_Y}{g_1^2} 
= \frac{1}{g_1^2}\left(\mhu-\mhd 
  + \sum_\text{gen}\left(\msq-2\msu+\msd-\msl+\mse\right)\right).
\end{equation}
Finally, we look for RGIs constructed from both scalar masses and gaugino 
masses. Note that the gaugino-mass $\beta$-functions can be rewritten as:
\begin{equation}
\beta(M_a^2) = 4b_ag_a^2M_a^2 \qquad (a=1,2,3).
\end{equation}
Combining the gaugino masses and scalar masses, we have fifteen 
$\beta$-functions with seven terms to eliminate, so we can construct eight RGIs
by taking linear combinations of the squared gaugino masses and the scalar 
masses. Five of them can be made from the scalar masses alone, so there must be
three new RGIs. In accordance with \cite{Carena1}, we take them to be
\begin{align}
I_{M_1} &\equiv M_1^2 - \frac{33}{8}\left(\msda-\msua-\msea\right), \\
I_{M_2} &\equiv M_2^2 + \frac{1}{24}\left(9\left(\msda-\msua\right) 
                + 16\msla-\msea\right), \\
I_{M_3} &\equiv M_3^2 - \frac{3}{16}\left(5\msda+\msua-\msea\right).
\end{align}
These complete the list of independent one-loop RGIs for the MSSM.

\end{document}